\documentclass[twocolumn]{emulateapj}

\usepackage{multirow}
\usepackage{graphicx}
\usepackage{amsmath, amsthm, amssymb}
\usepackage{color}
\usepackage{rotating}
\usepackage{chngcntr}
\usepackage{float}

\newcommand{\beq}{\begin{equation}}
\newcommand{\eeq}{\end{equation}}
\newcommand{\beqa}{\begin{eqnarray}}
\newcommand{\eeqa}{\end{eqnarray}}

\countdef\decade=200
\advance\decade by \year
\countdef\hours=201
\advance\hours by \time
\divide\hours by 60
\countdef\mins=202
\advance\mins by \hours
\multiply\mins by 60
\multiply\hours by 100
\countdef\miltime=203
\advance\miltime by \hours
\advance\miltime by \time
\advance\miltime by -\mins

\begin{document}
\title{Kiloparsec-Scale Simulations of Star Formation in Disk Galaxies. IV. \\
 Regulation of Galactic Star Formation Rates by Stellar Feedback}
\author{Michael J. Butler}
\affil{Max Planck Institute for Astronomy,  Konigstuhl 17, 69117 Heidelberg, Germany}
\author{Jonathan C. Tan}
\affil{Departments of Astronomy \& Physics, University of Florida, Gainesville, FL 32611, USA\\ \& National Astronomical Observatory, Mitaka, Tokyo 181-8588, Japan}
\author{Romain Teyssier} 
\affil{Institute for Computational Science, University of Zurich, 8049 Zurich, Switzerland}
\author{Joakim Rosdahl}
\affil{Leiden Observatory, Leiden University, PO Box 9513, NL-2300 RA Leiden, the Netherlands}
\affil{Univ Lyon, Univ Lyon1, Ens de Lyon, CNRS, Centre de Recherche Astrophysique de Lyon UMR5574, F-69230, Saint-Genis-Laval, France}
\author{Sven Van Loo}
\affil{School of Physics and Astronomy, University of Leeds, Leeds LS2 9JT, UK}
\author{Sarah Nickerson}
\affil{Institute for Computational Science, University of Zurich, 8049 Zurich, Switzerland}

\begin{abstract} 
Star formation from the interstellar medium of galactic disks is a
basic process controlling the evolution of galaxies. 
Understanding the star formation rate in a local patch of a disk with
a given gas mass is thus an important challenge for theoretical
models. Here we simulate a kiloparsec region of a disk, following the
evolution of self-gravitating molecular clouds down to subparsec
scales, as they form stars that then inject feedback energy by
dissociating and ionizing UV photons and supernova explosions. We
assess the relative importance of each feedback mechanism. We find
that $\rm H_2$-dissociating feedback results in the largest absolute
reduction in star formation compared to the run with no feedback.
Subsequently adding photoionization feedback produces a more modest
reduction. Our fiducial models that combine all three feedback
mechanisms yield, without fine-tuning, star formation rates that are
in excellent agreement with observations, with $\rm H_2$-dissociating
photons playing a crucial role. Models that only include supernova
feedback---a common method in galaxy evolution simulations---settle to
similar star formation rates, but with very different temperature and
chemical states of the gas, and with very different spatial
distributions of young stars.
%

\end{abstract}

\section{Introduction}
The formation of stars from the interstellar medium (ISM) of galactic
disks is one of the basic processes that controls the evolution of
galaxies. Most stellar populations are built-up in this way, including
those in typical Milky Way-like disk galaxies, as well as those formed
in starbursts due to galactic interactions that drive large quantities
of the ISM gas to more compact circumnuclear disks. Understanding the
star formation rate (SFR) per unit area, $\Sigma_{\rm SFR}$, from a
given galactic disk system as characterized by its total gas mass per
unit area, $\Sigma_g$, is thus a basic challenge to theoretical models
of star formation and galaxy evolution. Empirically, there is a
well-established correlation of $\Sigma_{\rm SFR} =
(6.3\pm1.8)\times10^{-3} (\Sigma_g/10\:M_\odot\:{\rm
  pc}^{-2})^{1.4\pm0.15}\:M_\odot\:{\rm yr}^{-1}\:{\rm kpc}^{-2}$ 
  based on global averages of disk systems (Kennicutt
1998). Considering only the mass surface density of molecular gas,
$\Sigma_{\rm H2}$, and averaging on smaller $\sim$kpc scales, a
linear relation $\Sigma_{\rm SFR} = (5.3\pm0.3)\times10^{-3}
(\Sigma_{\rm H2}/10\:M_\odot\:{\rm pc}^{-2})\:M_\odot\:{\rm
  yr}^{-1}\:{\rm kpc}^{-2}$ has been derived (Leroy et al. 2008;
Bigiel et al. 2008). These SFRs are of relatively low efficiency: only
about 4\% percent of the total gas content in a given annulus is
turned into stars per local orbital time (e.g., Leroy et al. 2008; Tan
2010). Zooming in to the scales where stars actually form, the
majority are born in very localized, $\sim$parsec-scale clumps, i.e.,
proto star clusters, within turbulent, magnetized giant molecular
clouds (GMCs), which are themselves $\sim$10 to 100~pc in size
(McKee \& Ostriker 2007). The inefficiency of star formation continues
down to these scales: only a few percent of GMC and clump gas forms
stars per local free-fall time (Krumholz \& Tan 2007; Da Rio et
al. 2014).

The physics of the star formation process is expected to involve a
competition between gravitational collapse and various forms of
pressure support in the GMCs and their clumps (e.g., thermal,
turbulent, magnetic, radiation). However, thermal pressure is
relatively unimportant at the $\lesssim20$~K temperatures of most GMC
material. The inefficiency of observed SFRs in GMCs implies that some
combination of turbulence, magnetic fields and feedback is playing a
crucial role in regulating SFRs, making the rates much smaller than
those that would result from unsupported, free-fall collapse of the
clouds. However, the relative importance of these inhibiting factors
remains very uncertain. On the theoretical side, this is because
numerical simulations of the ISM and star formation are challenging
due to the large range of scales that must be followed, the uncertain
choices that need to be made when specifying initial conditions, and
the wide variety of physical processes that must be considered.

In this paper we present numerical simulations that explore star
formation in galactic disks, focussing on the role of feedback in
regulating the ISM and star formation activity. This is the fourth in
a series of papers describing 3D simulations of a 1~kpc$^{2}$ patch of
a galactic disk extracted from the global galaxy simulation of Tasker
\& Tan (2009). In Van Loo, Butler \& Tan (2013, Paper I), we followed
the patch for 10 Myr down to a resolution of 0.5~pc.  Star formation
was included, with star particles with a minimum mass of
100~$M_{\odot}$ forming in gas above a density threshold of $n_{\rm
  H,*}>10^5\:{\rm cm^{-3}}$ at a star formation efficiency per local
free-fall time of $\epsilon_{\rm ff} = 0.02$.  No feedback processes
aside from a constant FUV field (of strength $4~G_{0}$, where
  $1\:G_0$ is the standard Habing (1968) intensity) were included,
and after 10 Myr we observed star formation rate surface densities
$\sim 100$ times higher than those in galaxies of comparable gas mass
surface density (e.g., Bigiel et al. 2008). In Paper I we speculated
that this was due to a lack of magnetic fields and/or local feedback
from young stars.

In Van Loo, Tan \& Falle (2015, Paper II) we explored the effects of
magnetic fields of varying strength on the same kpc-scale patch of the
Tasker \& Tan (2009) galactic disk.  Here we found that magnetic
fields suppressed the overall star formation rates by up to a factor
of two; however, this result was strongly influenced by the presence
of a magnetically supercritical starburst region.  In other regions,
larger suppression rates were observed.

In Butler, Tan \& Van Loo (2015, Paper III), we followed the kpc-scale
patch down to much higher resolution ($\sim 0.1$~pc) for 4 Myr.  Star
formation was included with the same recipe as in Papers I and II, but
now with a higher density threshold of $n_{\rm H,*}>10^6\:{\rm
  cm^{-3}}$ and a minimum star particle mass of 10 $M_{\odot}$.  As in
Paper I, we did not include magnetic fields or feedback from star
particles.  Enabled by our higher resolution, we explored the
structural, kinematic and dynamical properties of the large filaments
and clumps and compared them, when possible, to observations of
Infrared Dark Clouds (IRDCs) and long molecular filaments that have
been identified in the Galactic plane.  We found that by many metrics,
including too high dense ($\Sigma > 1.0$~$\rm g$~$\rm cm^{-2}$) gas
mass fractions, too high mass per unit length dispersion and velocity
gradients along the filaments, and too high velocity dispersion for a
given mass per unit length, our simulated filaments and clumps differ
greatly compared to observed clouds.  We therefore concluded that
IRDCs do not form from global fast collapse of GMCs, but rather we
expect them to be strongly influenced by dynamically important
magnetic fields.

The important features that distinguish the simulations of this
current paper (Paper IV) from previous works are: (1) the inclusion of
dissociating and ionizing feedback, in combination with supernova
feedback; (2) the adopted simulation set-up that is inherited from a
shearing, global disk simulation; (3) the high dynamic range from 1
kiloparsec down to 0.5 parsecs; (4) the high, realistic density
threshold for initiating star formation of $n_{\rm H,*}>10^5\:{\rm
  cm^{-3}}$ and the requirement that this gas be mostly molecular; (5)
the use of a star formation efficiency per local free-fall time of
2\%. These features of the star formation sub-grid model are motivated
by observational studies of star-forming molecular clouds (e.g.,
Zuckerman \& Evans 1974; Krumholz \& Tan 2007; Tan et al. 2014; Da Rio
et al. 2014). They are an approximate attempt to account for the
regulation of star formation rates by a combination of turbulence and
magnetic fields operating on sub-grid scales.

By comparison, Walch et al. (2015) presented simulations of a 500~pc
sized patch of a non-shearing disk with fiducial resolution of about
4~pc, studying only localized feedback from supernova, although in a
background UV radiation field. Kim et al. (2013) simulated a shearing
512~pc sized patch of disk down to 2~pc resolution to study star
formation from ``dense'' gas ($n_{\rm H,*}>200\:{\rm cm^{-3}}$) and
subsequent momentum injection from SN feedback.  Kim \& Ostriker
(2015) 
included the effects of magnetic fields, finding only modest
reductions in SFRs of ~25\%. On larger scales, Hopkins et al. (2012)
have presented smooth particle hydrodynamics (SPH) simulations of
galaxies with up to sub-pc resolution in which stars form when $n_{\rm
  H,*}>1000\:{\rm cm^{-3}}$ and including momentum feedback
representing stellar winds, radiation pressure and supernovae, and an
approximate Str\"omgren-type model for ionization feedback. However,
the SPH method has been criticized for being very diffusive and for
having difficulties of accurately modeling the multi-phase ISM that is
expected in regions affected by strong feedback (Agertz et
al. 2007). Agertz \& Kratsov (2015) have simulated a global disk
galaxy with a cosmological zoom-in grid-based AMR simulation that
includes energy and momentum feedback from winds, radiation pressure
and supernova feedback, but with resolution of only 75~pc. The
treatment of radiative transfer in the above simulations has generally
been very approximate in comparison to the moment based method we
utilize here. Rosdahl et al. (2015) have presented global galaxy
simulations including photo-ionization, radiation pressure and
supernova feedback, but with minimum resolutions of about 20~pc, still
too small to resolve GMCs.  Photodissociation feedback with a moment
based method of radiative transfer has not been included in any of the
above simulations.

We note that after submission of the first version of our paper,
  the study of Peters et al. (2017) has been published, which carried
  out similar simulations on a 0.5 kpc by 0.5 kpc non-shearing disk
  patch at 4 pc resolution that model star formation with a sink
  particle method and include dissociating feedback, in addition to
  photoionization, stellar winds and supernovae. One of the main
  conclusions of the Peters et al. study is that FUV dissociating
  feedback is needed to produce SFRs, i.e., gas depletion times, in
  agreement with observed galactic disks.

\section{Methods}

\begin{figure*}
\begin{center}$
\begin{array}{c} 
\includegraphics[width=6.5in]{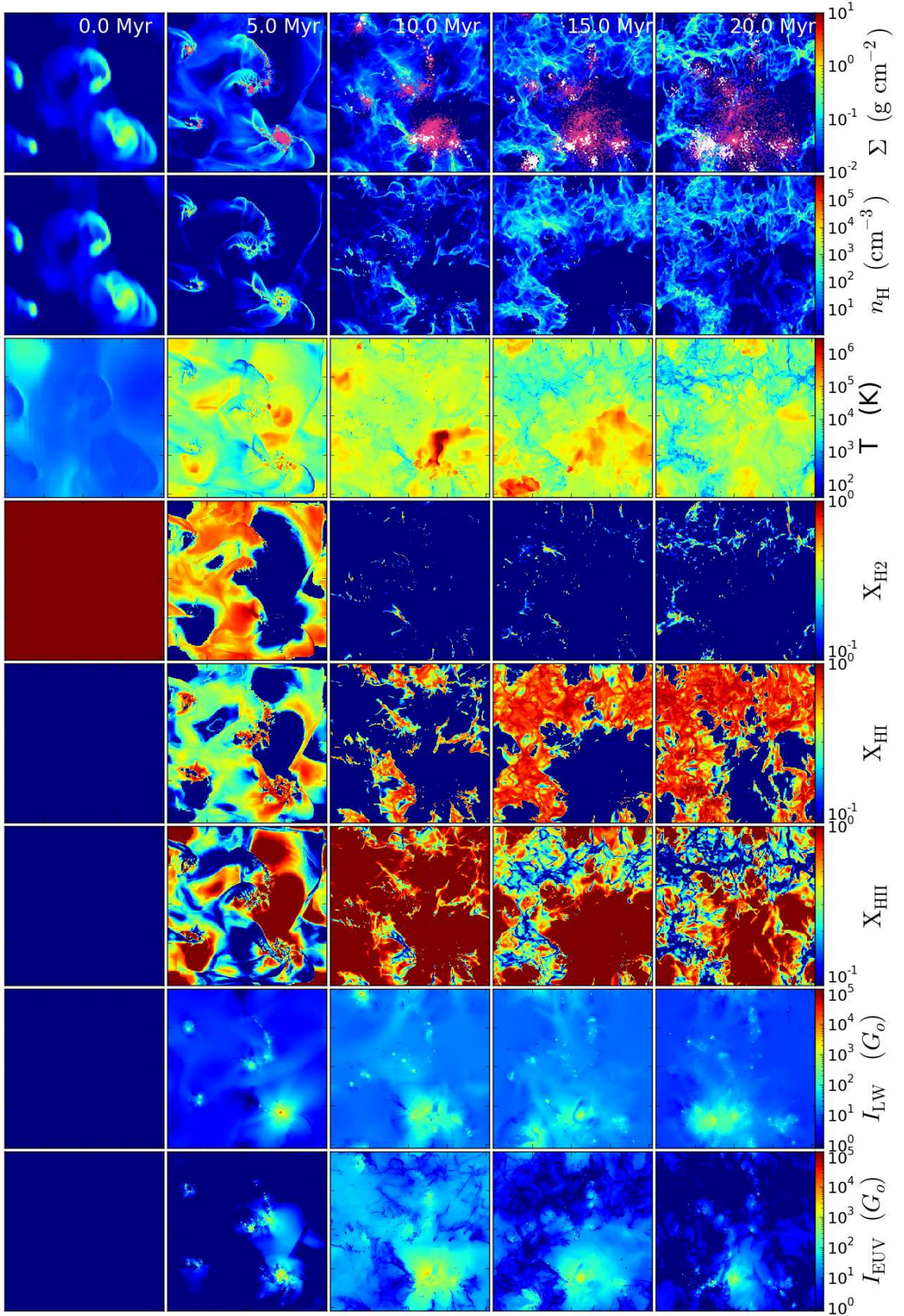} 
\end{array}$
\end{center}
\caption{\scriptsize
Evolution of the galactic interstellar medium for the fiducial
simulation run that includes dissociating, ionizing and supernova
feedback (Run LW+EUV+SN), as viewed projected along the $z$-axis from
above the disk (i.e., a top-down view, with each panel showing the
full 1~kpc sided disk patch, with box center at 4.25~kpc distance from
the galactic center, which is located to the left direction) at times
0, 5, 10, 15 and 20 Myr (left to right columns). Top row: total gas
mass surface density, $\Sigma_g$, along with young stars formed before
$5\:$~Myr (red points) and after $5\:$~Myr (white points). 2nd row:
mass-weighted hydrogen number density $n_{\rm H}$. 3rd row:
mass-weighted temperature, $T$. 4th, 5th and 6th rows: mass-weighted
molecular hydrogen fraction, $X_{\rm H2}$, atomic hydrogen fraction,
$X_{\rm HI}$, and ionized hydrogen fraction, $X_{\rm HII}$. 7th row:
mass-weighted Lyman-Werner (i.e., $\rm H_2$ dissociating) radiation
intensity $I_{\rm LW}$, which tends to emphasize the radiation field
that is having the main destructive effect on dense gas clouds. Bottom
row: mass-weighted hydrogen ionizing intensity $I_{\rm EUV}$.  }\label{fig:f1}
\end{figure*}

\begin{figure*}
\begin{center}$
\begin{array}{c} 
\includegraphics[width=6.7in]{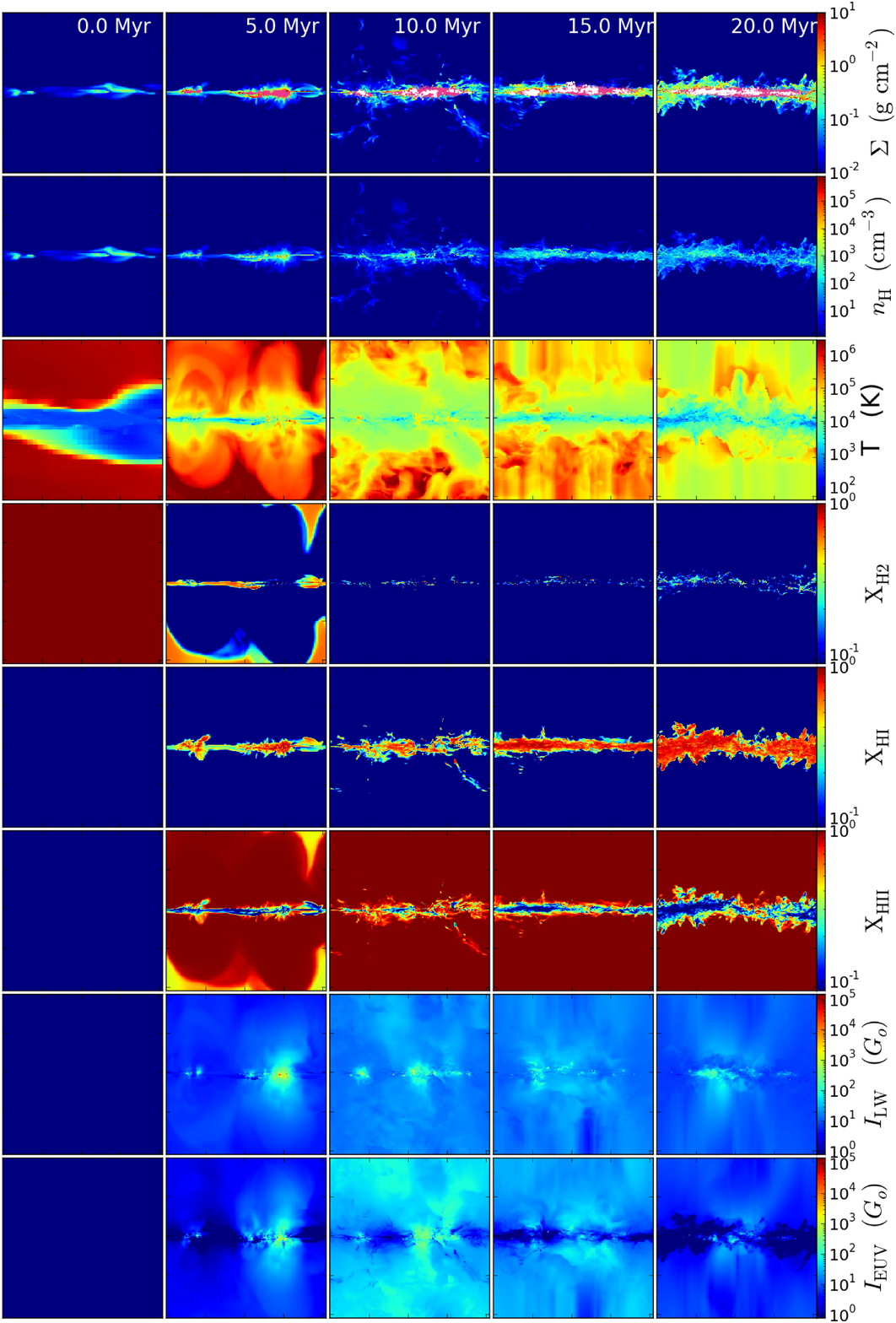} 
\end{array}$
\end{center}
\caption{
As Fig.~\ref{fig:f1}, but projected along the $y$-axis, i.e., an in-plane view,
for Run LW+EUV+SN.  }\label{fig:f2}
\end{figure*}

Our approach is to focus on a kiloparsec-sized patch of a typical
galactic disk, extended vertically also by a kiloparsec (i.e., to
500~pc either side of the disk midplane), with a flat, i.e., constant,
rotation curve of 200~$\rm km\:s^{-1}$ circular velocity. The center
of the patch is at a galactocentric radius of 4.25~kpc. This kpc-sized
region is large enough to be sensitive to the main effects of galactic
dynamics, including the shear flow due to differential rotation and
the vertical structure of the disk. At the same time the region is
small enough that it is practical to have resolution down to
0.5-parsec scales that can resolve GMCs. For initial conditions,  following Papers I, II \& III, we adopt a structured ISM with
average gas content of $\Sigma_g=17\:M_\odot\:{\rm pc}^{-2}$ that was
extracted from a lower, 8~pc, resolution global galaxy simulation
(Tasker \& Tan 2009 [TT09]), which already formed a population of
self-gravitating GMCs evolved to point of quasi statistical
equilibrium via shear-driven collisions and interactions. These
initial conditions are shown in the left columns of
Figures~\ref{fig:f1}
and \ref{fig:f2},
which display top-down and in-plane views of the simulation volume.
We note that the TT09 simulation did not include star formation
  or feedback and so the initial conditions here have only been
  sculpted by the processes of gravitational instability and cloud
  collisions in a shearing disk. As discussed by TT09 and Paper I, the
  GMCs have properties such as masses and virial parameters that are
  quite similar to observed GMCs, but with one main difference being
  that they have mass surface densities about a factor of two greater
  than local Milky Way GMCs. Partly for this reason, when these
  initial conditions are adopted in models that do include star
  formation and feedback there is an initial transient burst of star
  formation that we regard as unrealistic.  Thus to mitigate its
  effects we carry out evolution with an ``initial feedback model,''
  discussed below, in a first phase from 0 to 5 Myr, and then adopt
  the conditions at 5~Myr as our starting point for investigating the
  different effects of feedback implementation.

We note also that our simulation set-up, again following Papers
  I, II \& III, is not that of a shearing box, but rather is
  transformed into the rest frame of the center of the box at a
  galactocentric distance of 4.25~kpc. The gas in the disk patch
  inherits shearing motions that were present in the global disk (and
  GMCs also inherit a random velocity dispersion of $\sim10\:{\rm
    km\:s}^{-1}$), but these are not maintained during our
  simulation. For this reason, we limit the evolution to a period of
  20~Myr, i.e., about one flow crossing time at the maximum shear
  velocity. The boundary conditions of the volume are set to be
  outflow in the vertical directions and periodic in the disk plane.


Using the adaptive mesh refinement (AMR) radiation-hydrodynamics code
RAMSES-RT 
(Rosdahl et al. 2013), we have included several physical processes in the simulations,
especially: heating and cooling of molecular, atomic and ionized gas;
star formation from dense ($n_{\rm H}>10^{5}\:{\rm cm^{-3}}$),
molecular ($X_{\rm H2}>0.9$) gas, i.e., the creation of ``star
particles'' that each represents a small cluster of stars of
100~$M_\odot$, with average feedback properties assessed by assuming a
standard Chabrier (2003) stellar initial mass function (IMF); and
several feedback mechanisms from these newly-formed stars.  In
particular, we explore the effects of three types of stellar feedback:
H$_{2}$ dissociation by Lyman-Werner (LW) band ultraviolet photons,
photoionzation of hydrogen by extreme ultraviolet (EUV) photons that
have $>13.6$~eV energies, and supernovae, i.e., the explosive death of
massive stars that occurs 3~Myr after their formation leading to
release of $10^{51}$~ergs of kinetic energy. The radiation from the
stars is propagated on the AMR grid using a first-order Godunov
method, using the moment-based M1 approximation for the Eddington
Tensor. The primary advantage of this method is that it is independent
of the number of radiation sources, allowing us to study the effects
of feedback from many thousands of star particles.

As discussed above, the same initial conditions have also been
simulated previously with two other AMR simulation codes: {\it Enzo}
(Bryan et al. 2014) for the case of pure hydrodynamics with no direct
stellar feedback, but including the effect of a diffuse background FUV
radiation field (Van Loo, Butler \& Tan 2013; Butler, Tan \& Van Loo
2015); and {\it MG} (Falle et al. 2012) for the case of
magnetohydrodynamics (MHD), but with no direct feedback (Van Loo, Tan
\& Falle 2015). Thus we are able to compare our results with those of
these previous simulations to examine the relative importance of
magnetic fields and feedback.

The simulations contain four levels of AMR on top of the root grid
resolution of 7.8 pc (i.e., $128^3$ over the kpc$^3$ volume), reaching
a minimum cell size of 0.49 pc at the finest level (i.e., equivalent
to the resolution of a 2048$^3$ uniform grid simulation).  Cells are refined when the total mass within exceeds 800 M$_{\odot}$. For example, on refinement to the finest level, this corresponds to a density in a 1~pc sized cell of $n_{\rm H}=2.3\times10^4\:{\rm cm^{-3}}$, which is insufficient to resolve the Jeans length for temperatures that may be as low as $\sim 10\:$K. As with previous papers in this series, the primary goal is not to be able to properly resolve the details of gravitational fragmentation in any particular GMC, but rather to have an approximate estimate of the locations and mass fractions of gas that reach densities of $\sim10^5\:{\rm cm}^{-3}$ that are necessary for star formation.

We evolve the simulations for 20~Myr, first with an ``initial feedback
model'' until 5~Myr, and then with a variety of different feedback
implementations from 5 to 20~Myr. As described below, the goal of the
initial feedback model, which has reduced FUV feedback, is to mitigate
the effects of the initial transient burst of star formation that
depends on adopted initial conditions. 


In the following subsections, we describe the main physical processes
that have been included in the simulation code.

\subsection{Star Formation}

We utilize a sub-grid model for star formation in which gas is
converted, stochastically, into star particles at a fixed efficiency
$\epsilon_{\rm ff}=0.02$ per free-fall time, if a cell exceeds a given
threshold density, $n_{\rm H,*}=10^5\:{\rm cm^{-3}}$ and molecular
hydrogen mass fraction of $X_{\rm H2,*}=0.9$. This choice of
$\epsilon_{\rm ff}$ is motivated by the empirical results of Krumholz
\& Tan (2007) and is the same value used in the simulations of Van Loo
et al. (2015; 2013). The choice of $n_{\rm H,*}$ is also motivated by
observations of local star-forming clouds, especially Infrared Dark
Clouds (IRDCs), which can achieve these densities, while remaining
mostly starless (Butler \& Tan 2012; Tan et al. 2014). The choice of
$n_{\rm H,*}$ is also the same as that used in the simulations of Van
Loo et al. (2013; 2015). The value of $X_{\rm H2,*}$ is motivated by
the fact that, empirically, star formation is seen to occur
exclusively in molecular clouds. Physically, the reason for this may
be that this allows cooling to low temperatures of $\sim10\:$K, helped
by CO line cooling, which then allows dense, supersonically turbulent
gas clumps to form. Their relatively low ionization fractions and
turbulent conditions may be crucial for mediating the loss of magnetic
flux support from the gas, e.g., by turbulent reconnection (Lazarian
\& Vishniac 1999; Eyink et al. 2011).


Star formation is then modeled by creation of ``star particles'' that
are intended to represent a small cluster of young stars. On their
creation, the equivalent mass is removed instantaneously from the gas
in the relevant cell. The star particles only feel gravitational
forces from the surrounding gas and the background galactic
potential. Also, they do not accrete any material once formed. We
adopt a minimum star particle mass of $100\:M_\odot$, with this choice
being motivated by the minimum gas mass that is present in a
star-forming 0.5~pc-sized cell of $400\:M_\odot$, given the choice of
$n_{\rm H,*}=10^5\:{\rm cm^{-3}}$. In typical simulation time steps,
the mass of new young stars that should form is much less than
$100\:M_\odot$, so a star particle is only formed with a probability
that yields an overall, time-averaged SFR equal to that set by the
choice of $\epsilon_{\rm ff}=0.02$. Thus all star particles in the
simulations are in fact born with masses of $100\:M_\odot$ and when
they are formed, the mass of gas in the natal cell undergoes at most a
$\sim25\%$ change.

\subsection{Stellar Evolution and Feedback}\label{S:methods_feedback}

The star particles are then sources of feedback on the surrounding
gas. The feedback properties are based on that expected from a stellar
population with initial masses distributed according to the Chabrier
(2003) IMF
with upper and lower mass limits of 120 $M_\odot$ and 0.1
$M_\odot$. In reality, one expects stochastic sampling of the upper
end of the IMF, but this effect is not included here. We anticipate
that it would make only minor differences due to the large number of
star particles that are formed in the simulations.  Thus we utilize
IMF-averaged stellar evolutionary tracks from Bruzual \& Charlot
(2003) that follow evolution from the zero age main sequence to later
stages (see Rosdahl et al. 2015 for more details).

In this study, our primary focus is to investigate the thermal and
chemical feedback due to UV radiation. Thus the FUV radiation field is
constructed, which induces photoelectric heating feedback. It also
leads to dissociation of H$_2$. The EUV radiation field leads to
ionization of H, with the ionized gas heated to near $10^4\:$K. The
elevated thermal pressure of these HII regions is another form of
feedback. The radiative transfer methods necessary to model this
feedback are described below, while the types of radiation feedback
are summarized in Table~\ref{tab:rt}.

Note, in this work, we have not modeled the effects of radiation
pressure feedback, which is deferred to a future study (Rosdahl et
al., in prep.). Nor have we modeled the momentum input from stellar
wind feedback, which is difficult to resolve on the scales followed
here, especially the effects of overlapping wind bubbles from
individual stars, which will lead to a reduction in the overall
momentum injection due to oppositely directed flows.




Supernova feedback is modelled with a single injection from each star
particle (of total mass $m_*$) into its host cell 3~Myr after the
particle is formed of ejecta mass $m_{\rm ej} = \eta_{\rm SN} m_{*}$
and thermal energy $\epsilon_{\rm SN} = 10 \eta_{\rm SN} \times
10^{51} {\rm erg} (m_{*}/100\:M_{\odot})$.  We choose a value of
$\eta_{\rm SN} = 0.1$, resulting in each star particle injecting the
energy and mass expected from a single supernova progenitor star of
initial mass of approximately 10 to 20~$M_{\odot}$.  It is worth
noting that, thanks to our high mass resolution, we are effectively
modelling each individual supernovae explosion, one star at a time.
Following the supernova event, cooling is delayed by a dissipation
timescale $t_{\rm diss}$, which we have set to 50,000~yr, given
typical densities on the scales of our minimum resolution of
0.5~pc. In practice, this is implemented by turning off cooling in a
cell (and adjacent cells) that contains a star particle that underwent
a supernova explosion in the last 50,000~yr (see Teyssier et al. 2013;
Rosdahl et al. 2015).

To enable the rapid development of a realistic ISM structure that is
approximately similar to that in the inner Milky Way disk, e.g., at
$\sim4\:$kpc galactocentric radius, we first run the simulation from 0
to 5 Myr with an ``initial feedback model'' that includes FUV, EUV and
SN feedback. However, because there is an initial transient burst of
star formation that occurs before feedback can begin to fully regulate
the ISM and because the FUV feedback from stars is quite long-lived
(i.e., $\sim10\:$Myr), we artificially lower the intensity of FUV
radiation from these initial phase star particles by a reduction
factor of 0.1, while keeping EUV and SN feedback as normal. The net
result of this model is that by 5 Myr, the FUV intensity from the
initial phase stars is at a median level of 5.3~$G_0$ within the
100~pc thick disk midplane region, where $1\:G_0$ is the standard
Habing (1968) intensity. This value is similar to that of the Milky
Way at galactocentric radii of $\sim4\:$kpc (Wolfire et al. 2003). The
main focus of our work is then to investigate how implementing
different feedback methods from stars formed after 5~Myr affects ISM
and star formation properties from the period from 5 to 20~Myr. Note,
that the residual effects of feedback from the initial phase stars
continue after 5~Myr, but quite rapidly lose importance due to the
short timescales associated with massive star evolution.


\subsection{Radiative Transfer}

Details of the radiative transfer algorithm can be found in Rosdahl et
al. (2013).  Here we outline the main methods and approximations used
in this paper. For stability reasons, the radiative transfer time-step
scales inversely with the speed of light $c$.  To avoid very small
timesteps relative to those set by the hydrodynamics, we use the
``reduced speed of light approximation'' (Gnedin \& Abel 2001), and
limit the propagation of radiation to a global speed, $\bar{c}_{\rm
  red}\ll c$, where $c$ is the true speed of light.  In our
simulations we adopt the value of $\bar{c}_{\rm red} = 0.005 c$, i.e.,
1500~km~s$^{-1}$. We expect the effects of this choice to be
relatively minor (see Rosdahl et al. 2015 for explorations of the
effect of this choice on SFRs in larger-scale simulations).  It is
important to note that this approximation is only valid if the reduced
speed of light $\bar{c}_{\rm red}$ results in a light crossing time
that is short compared to smallest of the sound crossing time, the
recombination time and the advection time. We have checked {\it a
  posteriori} that our simulation indeed satisfies these requirements.
For the transport of radiation, as described in detail in Rosdahl \&
Teyssier (2015), we use a first order moment method, which tracks in
each cell the radiation energy density $E$ (zeroth moment, energy per
unit volume) and the bulk radiation flux vector {\bf F} (first moment,
energy per unit area and time). The system of moment equations is
closed with the M1 expression for the radiation pressure tensor, which
has the large numerical advantage that it can be evaluated purely from
the local quantities $E$ and {\bf F}.  A moment method treats the
radiation as a fluid, rather than tracking individual rays, giving the
very important advantage that we can treat a large number
of radiation sources. As demonstrated in detail in Rosdahl \& Teyssier
(2015), the sacrifice that comes with this is that rays from
individual sources are treated in an average sense in every volume
element, resulting in a radiation field which is not as accurate as
with ray-tracing, on scales of the source separation. In Rosdahl et
al. (2015), we demonstrate that qualitatively we still retain a
correct radiation field in a galaxy disk setup.



The two moments are the radiation energy density $E$ (erg $\rm
cm^{-3}$) and the radiation flux $\bf{F}$ (erg cm$^{-2} \rm
s^{-1}$). For the latter, we need to model the radiation pressure
$\bf{P}$ (erg cm$^{-3}$), for which we use the M1 closure. The set of
moment equations is solved separately for each radiation group, which
is defined for a specific frequency range.  For
this work, we model four radiation groups (see Table~\ref{tab:rt})
that determine the photodissociation of $\rm H_{2}$, the ionization of
$\rm HI$, the ionization of $\rm HeI$, and the ionization of $\rm
HeII$, respectively.

\begin{deluxetable*}{ccccccc}
\tablecolumns{7}
\tablewidth{0pt}
\tablecaption{Radiation Group Energy Intervals and Properties$^{a}$}
\tablehead{
\colhead{Photon Group} & \colhead{$\epsilon_{0}$} & \colhead{$\epsilon_{1}$} & \colhead{$\sigma_{\rm H2}$} & \colhead{$\sigma_{\rm HI}$} & \colhead{$\sigma_{\rm HeI} $} & \colhead{$\sigma_{\rm HeII} $} \\
\colhead{} & \colhead{eV} & \colhead{eV} & \colhead{$\rm cm^{2}$} & \colhead{$\rm cm^{2}$} & \colhead{$\rm cm^{2}$} & \colhead{$\rm cm^{2}$}
}

\startdata
LW &  12.00 & 13.60 & 3.000 $\times 10^{-22}$ & 0.000 & 0.000 & 0.000 \\
EUV$_{\rm HI}$ & 13.60 & 24.59 & 0.000 & 3.007 $\times 10^{-18}$ & 0.000 & 0.000 \\
EUV$_{\rm HeI}$ & 24.59 & 54.42 & 0.000 & 5.687 $\times 10^{-19}$ & 4.478 $\times 10^{-18}$ & 0.000 \\
EUV$_{\rm HeII}$ & 54.42 & $\infty$ & 0.000 & 7.889 $\times 10^{-20}$ & 1.197 $\times 10^{-18}$ & 1.055 $\times 10^{-18}$ \\
\enddata
\tablenotetext{a}{
Group intervals are
given in units of eV by $\epsilon_{0}$ and $\epsilon_{1}$.  The next
four columns show the dissociation (for $\rm H_{2}$) and ionization
cross sections for hydrogen and helium, and are derived every 5 coarse
time-steps from the stellar luminosity weighted SED model (Bruzual \&
Charlot 2003). These properties evolve over time as the stellar
populations age, varying by a few percent.
}
\label{tab:rt}
\end{deluxetable*}


The ionization fractions are defined as $X_{\rm HII} = n_{\rm
  HII}/n_{\rm H}, X_{\rm HeII} = n_{\rm HeII}/n_{\rm H}, X_{\rm HeIII}
= n_{\rm HeIII}/n_{\rm H}$.  While previous work using Ramses-RT at
much larger scales only considered the atomic and ionized species of
hydrogen and helium (e.g., Rosdahl et al. 2015), with our higher
resolution and denser, colder gas, it becomes important to model the
formation and destruction of molecules.  We introduce a
photo-dissociating radiation model for molecular hydrogen that is
included in the non-equilibrium chemistry solver.  H$_{2}$ is formed
on dust grains as described by Draine \& Bertoldi (1998), as well as
from the gas phase onto H$^{-}$, assuming equilibrium abundances for
H$^{-}$ (see Abel et al. 1997), the latter of which is only important
at low metallicities.

H$_{2}$ is dissociated either through collisions with HI (Dove \&
Mandy 1986; Abel et al. 1997), or through photodissociation by photons
in the Lyman-Werner band.  For the latter, we use the radiative
transfer algorithm of RAMSES-RT, without considering Doppler effects
and the associated relativistic corrections to the moment
equations. In other words, this amounts to ignoring the effects of
velocity gradients in the H$_2$ opacity calculation, while still
properly accounting for shielding of LW radiation due to the column
density of H$_{2}$ in dense molecular clouds.  These effects are also
ignored for the ionization of hydrogen and helium.

\subsection{Heating and Cooling}

The hydrogen and helium thermochemistry is calculated based on
absorption and emission of photons by hydrogen and helium species,
photoelectric heating due to electrons ejected from dust grains, as
well as equilibrium metal and dust cooling assuming solar metallicity.
The gas thermal energy density
is tracked in each cell along with the abundances of the species that
interact with the photons.  We utilize the same non-equilibrium
thermochemical network described in Rosdahl et al. (2015).  In this
network, HI, HeI, and HeII interact with photons through
photoionization, and HII, HeII, and HeIII through recombination.  In
this paper we have additionally considered the cooling and heating
contributions from the dissociation of molecular hydrogen $\rm H_{2}$
by photons in the Lyman-Werner band, as well as line emission from
molecular rotational and vibrational levels and fine-structure atomic
levels.  We also include free-free, free-bound, and two photon
cooling, as well as photoelectric heating via dust grains that scales
with the Lyman-Werner flux as in Forbes et al. (2016).

While we model non-equilibrium hydrogen and helium thermochemistry, a
non-equilibrium description of the thermochemistry of metal species is
beyond the scope of this work. Instead, we use tabulated metal
contributions to the cooling rates, including dust cooling, 
extracted from CLOUDY (Ferland et al. 2013), stored as a function of
the local temperature, and assumed to scale linearly with the gas
density and metallicity.



\section{Results}

\subsection{Effect of Feedback on Global Evolution of the Interstellar Medium}\label{S:ism}

We follow the evolution of the ISM in a series of galactic disk patch
simulations for 20~Myr. We will see that this is a long enough time
for the results of different feedback mechanisms to become readily
apparent and for the properties of the ISM, including its SFR, to
reach a quasi statistical equilibrium. As discussed in
\S\ref{S:methods_feedback}, to help enable the rapid development of a
realistic ISM structure that is approximately similar to that in the
inner Milky Way disk, e.g., at a distance of $\sim4\:$kpc from the
Galactic, we first run the simulation from 0 to 5 Myr with an
``initial feedback model'' that includes FUV (reduced by a factor of
0.1), EUV and SN feedback. 
The net result of this model is that by 5 Myr, the FUV intensity from
the ``initial phase'' stars is at a median level of 5.3~$G_0$ within
the 100~pc thick disk midplane region.

After 5 Myr, different cases of feedback are then investigated for the
newly formed stars in a sequence of model runs: (1) No feedback; (2)
only FUV feedback; (3) FUV+EUV feedback; (4) FUV+EUV+SN feedback (the
fiducial case); (5) only SN feedback. Note, that the feedback from the
initial phase stars continues after 5~Myr in all cases, but the
effects of their EUV and SN soon die out, while the impact of the
reduced FUV feedback from these stars also gradually declines. ISM
properties and star formation activity from 5 to 20 Myr and their
dependencies on how feedback is modeled are the focus of this paper.

Figures~\ref{fig:f1} and \ref{fig:f2} show images of the fiducial simulation results, i.e.,
FUV+EUV+SN feedback, after 5, 10, 15 and 20 Myr as viewed from above
(top-down view) and from the side (in-plane view) of the disk,
respectively. The quantities displayed are: the mass surface density
structure of the gas, $\Sigma_g$, together with locations of young
stars, distinguishing those that form after 5~Myr from those that
formed earlier; number density of H nuclei, $n_{\rm H}$ (this quantity
and the following are mass-weighted along the line of sight), gas
temperature, $T$, $\rm H_2$, $\rm H$ and $\rm H^+$ mass fractions,
$\rm H_2$ dissociating (Lyman-Werner) photon intensities, and
H-ionizing (EUV) photon intensities.

Figure~\ref{fig:f1}'s top-down view is the clearest way to visualize the
evolution of the several initial GMCs that are orbiting in the
galactic disk. With the velocity reference frame of the simulation set
equal to the orbital velocity of the center of the box, the shear flow
causes GMCs on the left, inner-galaxy side to tend to move upwards in
the figures, while clouds on the right, outer-galaxy side move
downwards. However, note that the clouds also inherit random motions,
$\sim 10\:{\rm km\:s^{-1}}$, from the global galaxy
simulation. Periodic boundary conditions are adopted for the sides,
while outflow boundary conditions are applied on the top and bottom
sides of the cube. Using AMR, finer resolution grids down to 0.5~pc
scales are applied (i.e., up to 4 levels of refinement), which follow
fragmentation of the GMCs into dense gas clumps.
\begin{figure*}
\begin{center}$
\begin{array}{c} 
\includegraphics[width=6.5in]{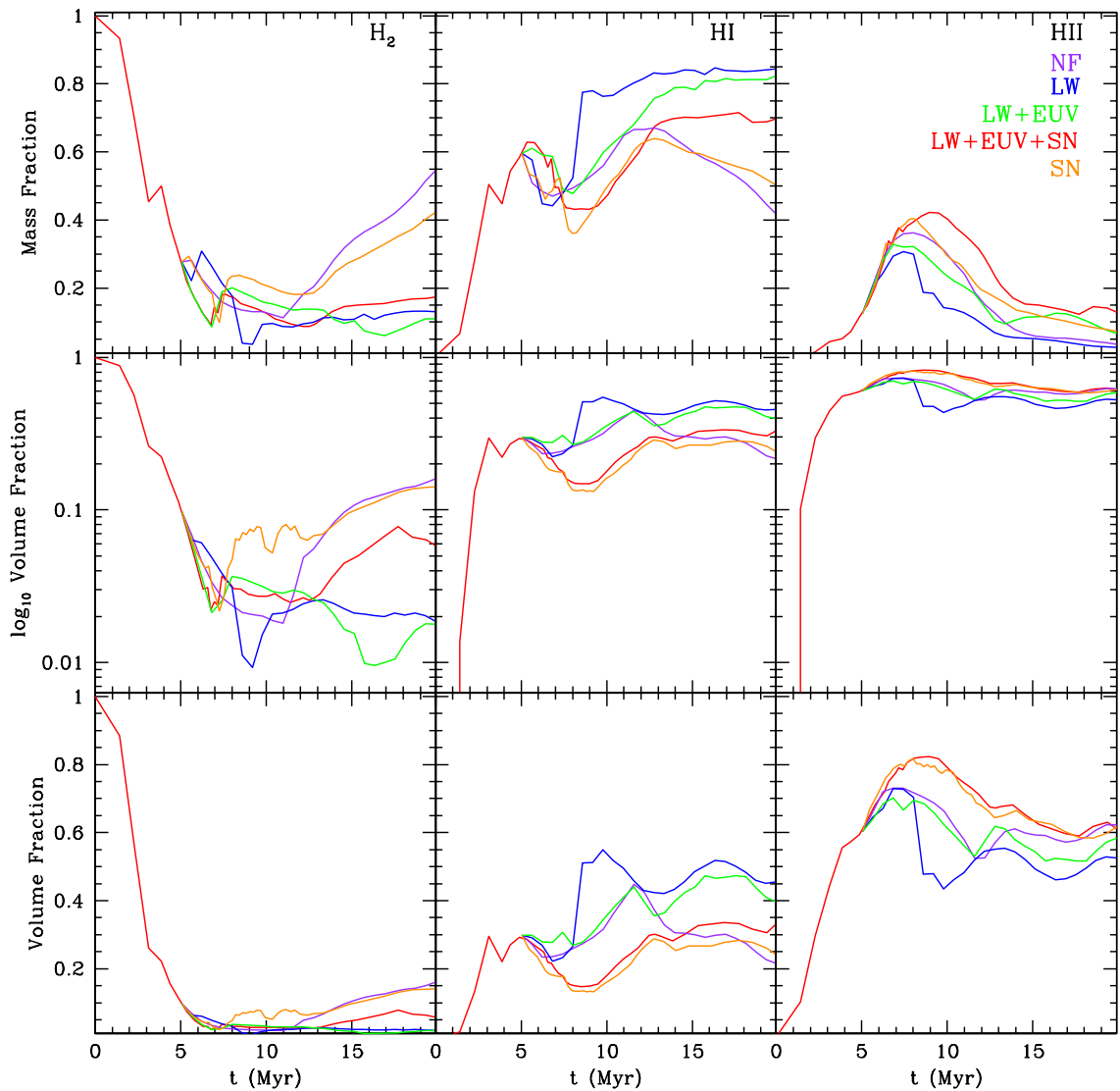} 
\end{array}$
\end{center}
\caption{
Time evolution of the hydrogen phase mass fractions (top row) and
volume fractions (middle and bottom rows) of the 100~pc thick disk
region for all simulations. The fiducial simulation (LW+EUV+SN)
achieves a quasi-steady state in these phase fractions during the
period from $\sim$10 to 20~Myr.}\label{fig:f3}
\end{figure*}

\begin{figure*}
\begin{center}$
\begin{array}{c} 
\includegraphics[width=6.7in]{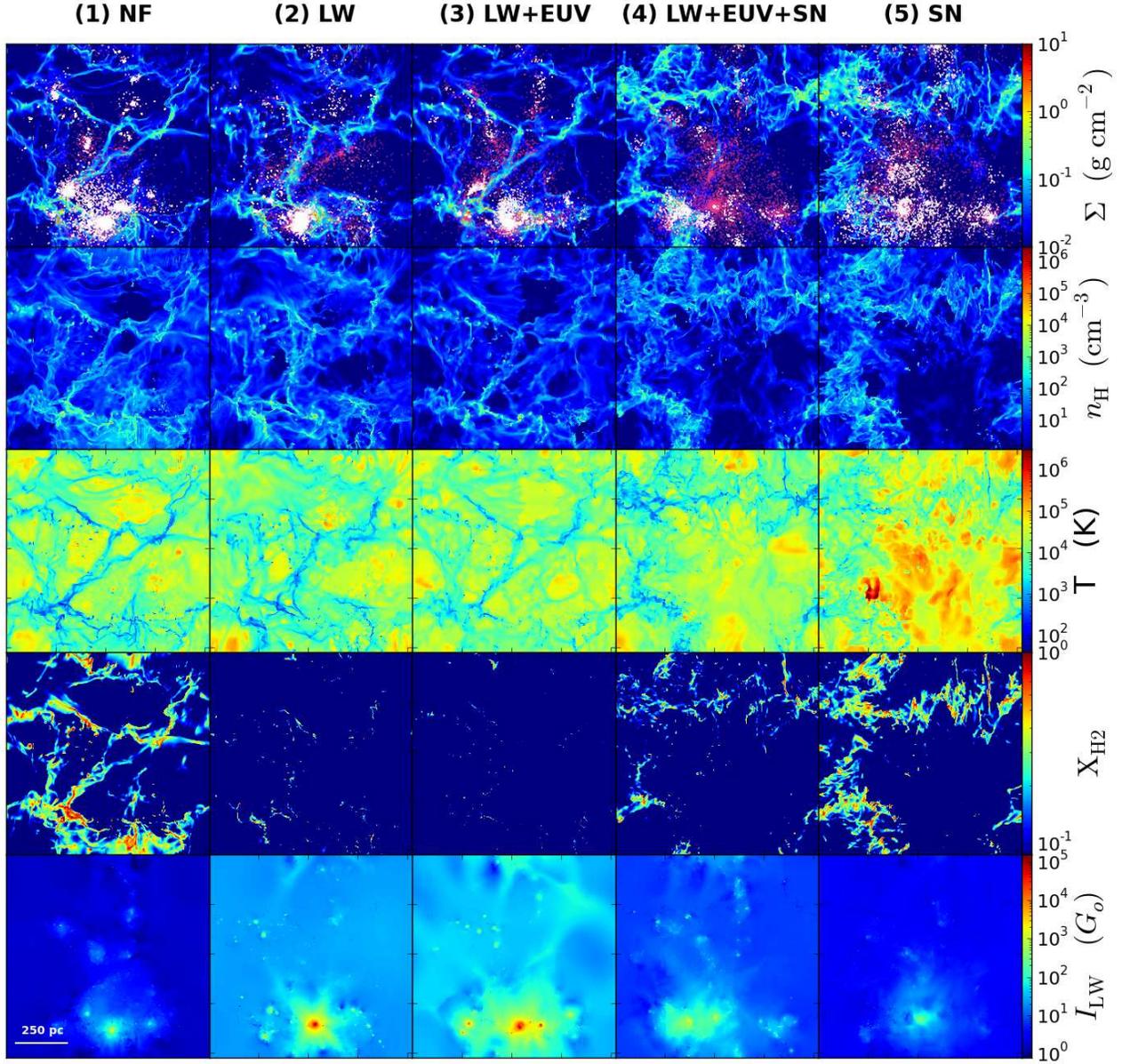} 
\end{array}$
\end{center}
\caption{ 
Comparison of feedback models. The kpc disk patch is shown in each
panel at $t=$20~Myr (top-down view). Columns left to right show: (1)
No Feedback (NF); (2) Only Dissociating Lyman-Werner Feedback (LW);
(3) Dissociating \& Ionizing Feedback (LW+EUV); (4) Dissociating,
Ionizing \& Supernova Feedback (LW+EUV+SN) (fiducial model); (5) Only
Supernova Feedback (SN). Top row: total gas mass surface density,
$\Sigma_g$, along with young stars formed before $5\:$~Myr (red
points) and after $5\:$~Myr (white points).
2nd row: mass-weighted H number density, $n_{\rm H}$; 3rd row:
mass-weighted temperature, $T$; 4th row: mass-weighted $\rm H_2$
fraction, $X_{\rm H2}$; bottom row: mass-weighted Lyman-Werner
radiation intensity, $I_{LW}$.}\label{fig:f4}
\end{figure*}

\begin{figure*}
\begin{center}$
\begin{array}{c} 
\includegraphics[width=6.7in]{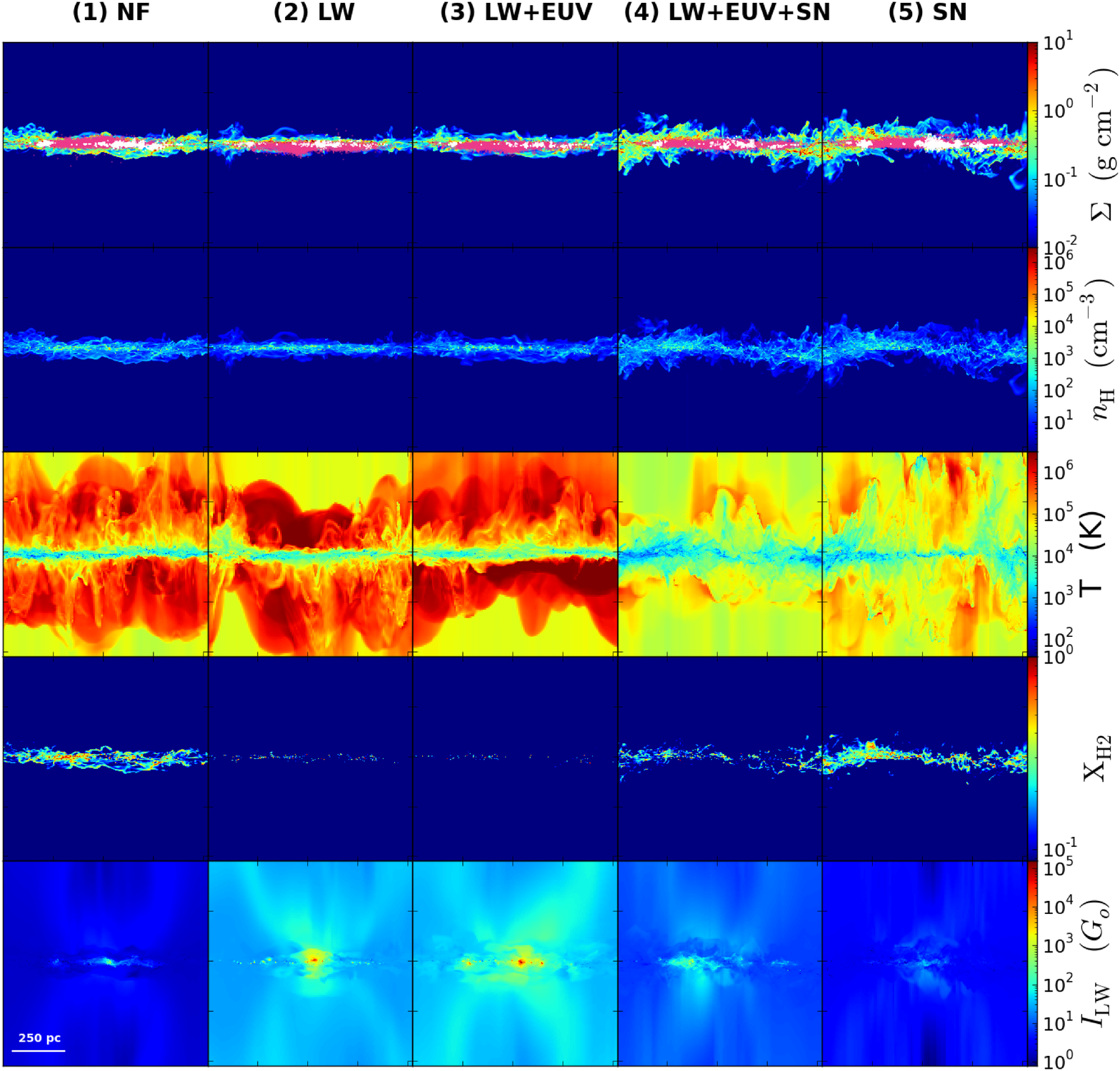} 
\end{array}$
\end{center}
\caption{ 
As Fig.~\ref{fig:f4}, but now viewed with an in-plane view,
projected along the $y$-axis.
}\label{fig:f5}
\end{figure*}

Above our adopted threshold density of $n_{\rm H,*}=10^5\:{\rm
  cm^{-3}}$
and for gas that is $>$90\% molecular, stars are formed with the
empirically motivated subgrid model that turns gas into star particles
at an efficiency of 2\% per local free-fall time (Krumholz \& Tan
2007), described in \S\ref{S:methods_feedback}. These star particles
then act as sources of dissociating, ionizing and, after a 3~Myr
delay, supernova feedback. The radiative feedback can be seen directly
in the figures showing the intensities of dissociating and ionizing
radiation, which then influence the temperature and chemical state of
the gas. The intensity of the UV Lyman-Werner radiation field varies
by orders of magnitude, being very high in regions of active star
formation. By 20~Myr the median FUV intensity in the 100~pc thick disk
midplane region is about 8.0~$G_0$.

In Figure~\ref{fig:f3} we show the time evolution of the mass and
volume fractions of the various ISM phases in the 100~pc thick disk
region for the different feedback simulations. For the fiducial
simulation at 10, 15 and 20~Myr, the ISM has overall mass fractions:
$X_{\rm H2}=0.124, 0.154, 0.174$, $X_{\rm HI}=0.474, 0.695, 0.699$ and
$X_{\rm HII}=0.402, 0.151, 0.127$, respectively. The equivalent volume
fractions are: $V_{\rm H2}=0.0273, 0.0498, 0.0571$, $V_{\rm HI}=0.183,
0.326, 0.336$ and $V_{\rm HII}=0.790, 0.624, 0.607$ (we note that only
a very small fraction of the total gas mass, $\lesssim 3\%$ is
expelled from the simulation box to distances $>500\:$pc). After a
burst of ionizing feedback, which raises $X_{\rm HII}$ to about 0.4,
these mass and volume fractions of the different phases approach
relatively constant values as the simulation evolves towards 20~Myr,
illustrating that by these metrics the ISM reaches a quasi-statistical
equilibrium.  We note that by 20~Myr the fiducial simulation has
  a ratio $X_{\rm H2}/(X_{\rm H2}+X_{\rm HI})\simeq 0.2$, which is 2
  to 3 times smaller than the average values inferred inside the solar
  circle by Koda et al. (2016). We expect this is due to the
  moderately elevated values of FUV intensity that are present, i.e.,
  equivalent to $8 G_0$, about double that in the inner Galaxy.

Comparing the results of the different feedback simulations in the
sequence from Runs 1 to 4 (NF, LW, LW+EUV, LW+EUV+SN), we see the $\rm
H_2$ mass fraction at 20~Myr being reduced to close to the
quasi-equilbrium value mostly by the introduction of LW feedback. On
its own, this LW feedback keeps most of the gas in the HI phase. With
introduction of EUV and SN, a more significant mass fraction of the
ionized phase is created. SN-only feedback leads to a much higher $\rm
H_2$ mass fraction.

The equivalent results to Figures~\ref{fig:f1} and \ref{fig:f2} for the simulations that
have different feedback implementation after 5 Myr are shown in the
Appendix, i.e., no feedback (Run NF), only
dissociating Lyman-Werner feedback (Run LW), dissociating plus
ionizing feedback (Run LW+EUV) and only supernova feedback (Run SN)
(Figures A1 - A8).
The choice of how feedback is implemented can have
a profound effect on the structure of the ISM. This is also
illustrated in Figures~\ref{fig:f4} and \ref{fig:f5}, which show the results of all the
above simulations side-by-side after 20~Myr of evolution.

The NF run retains denser clouds and the global ISM is generally
cooler, with heat input only from the declining effects of feedback
from the initial phase (0 to 5~Myr formed) stars. Adding LW feedback
results in much smaller $\rm H_2$ mass and volume fractions and a
generally warmer ISM. However, large scale distributions of dense gas
structures remain relatively unaffected compared the NF model. Adding
EUV feedback continues these trends. Long, $\gtrsim100\:$pc, cold,
dense filamentary structures are seen aligned with the galactic
plane. Finally, the fiducial model with LW+EUV+SN feedback leads to
more disruption in the distribution of the dense gas structures: e.g.,
filaments tend to be shorter and/or more irregular.  There are
enhanced $\rm H_2$ fractions in some regions due to sweeping up of
previously atomic gas. More strongly fluctuating gas potentials also
appear to lead to a greater dispersal of the stars that are born in
bound star clusters. SN-only feedback leads to much more disruption of
the initial molecular clouds, higher fractions of the ISM in the hot
($\gtrsim10^6\:$K) phase, but also larger fractions of the molecular
phase. The young stars are more dispersed in the SN-only feedback run.

\begin{figure*}
\begin{center}$
\begin{array}{l} 
\hspace{-0.5in}
\includegraphics[width=7.7in]{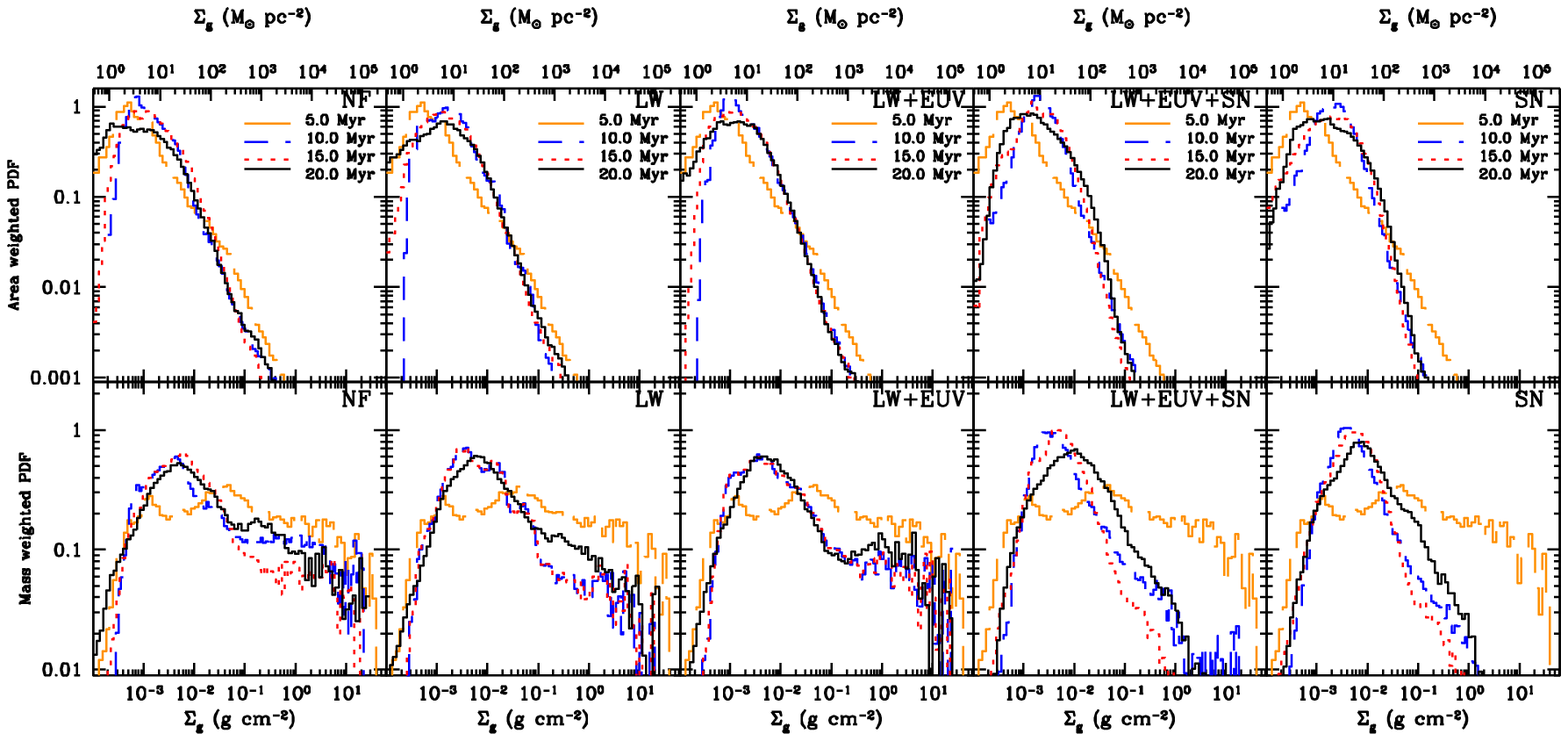} \\
\end{array}$
\end{center}
\vspace{-4in} 

\caption{
Time evolution from 5 to 20 Myr of the area-weighted (top row) and
mass-weighted (bottom row) gas mass surface density probability
distribution functions (as viewed from above the disk) for the (from
left to right columns): no feedback run, LW run, LW+EUV run, LW+EUV+SN
run, and SN-only run.
}\label{fig:f6}
\end{figure*}

\begin{figure*}
\begin{center}$
\hspace{-0.5in}
\includegraphics[width=7.5in]{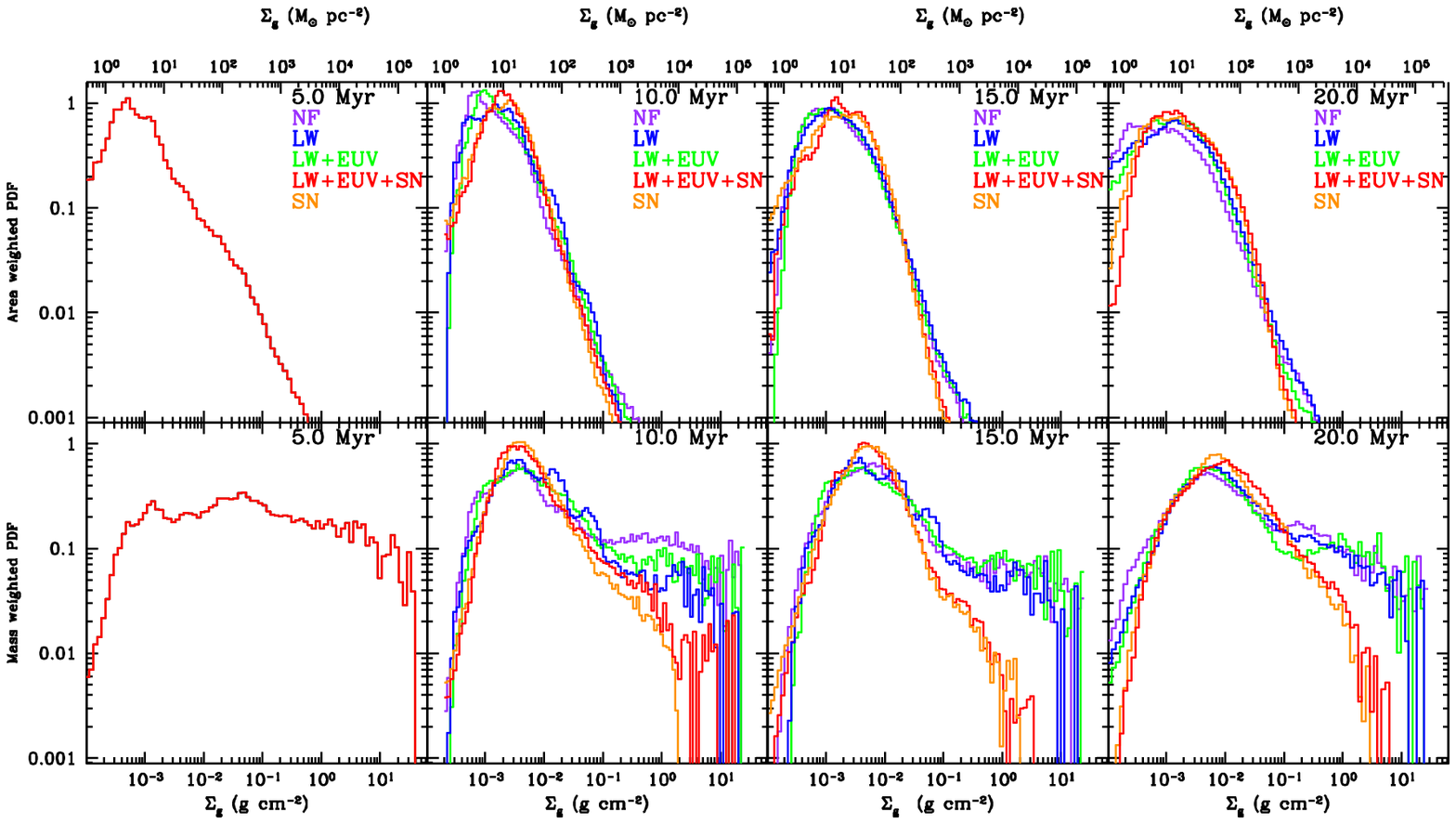}$
\end{center}
\vspace{-3.2in} 
\caption{
Area-weighted (top) and mass-weighted (bottom) gas mass surface
density probability distribution functions (as viewed from above the
disk) at 5.0, 10.0, 15.0, and 20.0 Myr for the no feedback run (purple),
LW run (blue), LW+EUV run (green), LW+EUV+SN run (red), and SN only
(orange).
%
}\label{fig:f7}
\end{figure*}

In Figures~\ref{fig:f6} and \ref{fig:f7}, we present comparisons of the area and
mass-weighted probability distribution functions (PDFs) of gas mass
surface densities, viewed from above the disk, that arise in the
different models. These are amongst the simplest statistical metrics
of the ISM, which can be compared to observed systems, especially as
we enter the era of full operation of the {\it Atacama Large Mm-submm
  Array} (ALMA) that has the ability to resolve $\sim$parsec scales in
nearby galaxies. Figure~\ref{fig:f6} shows that relatively large changes in the
PDFs occur from 5 to 10~Myr in all models and then subsequent
evolution is more gradual. The mass fraction at high $\Sigma$
conditions decreases, which is a consequence of both formation of
stars and disruption of gas clouds by feedback.

Figure~\ref{fig:f7} compares the different feedback models at
particular times in the simulations, starting from identical initial
conditions at 5~Myr. The rapid changes from 5 to 10~Myr are again
obvious. By 20~Myr, the area-weighted $\Sigma$ PDFs in the models with
feedback achieve distributions with a single main peak just below
$10\:M_\odot\:{\rm pc}^{-2}$. In the mass-weighted PDFs, the
high-$\Sigma$ component is more apparent and at 15 and 20~Myr we
notice quite similar distributions in the NF, LW and LW+EUV
models. However, there can be factors of a few differences in the
  amount of material with $\Sigma\sim 0.4\:{\rm g\:cm}^{-3}$, which is
  close to the value of individual 0.5~pc cells on reaching the
  threshold volume density for star formation of $n_{\rm
    H,*}=10^5\:{\rm cm}^{-3}$ for star formation. The effect of SN in
the LW+EUV+SN and SN-only runs is to destroy much of the highest
$\Sigma$ structures, i.e., $\gtrsim 1\:{\rm g\:cm^{-2}}$.
Another feature of the PDFs is that those of the LW+EUV+SN and SN-only
runs are quite similar to each other.

\begin{figure*}[ht!]
\begin{center}$
\begin{array}{c} 
\includegraphics[width=7.4in]{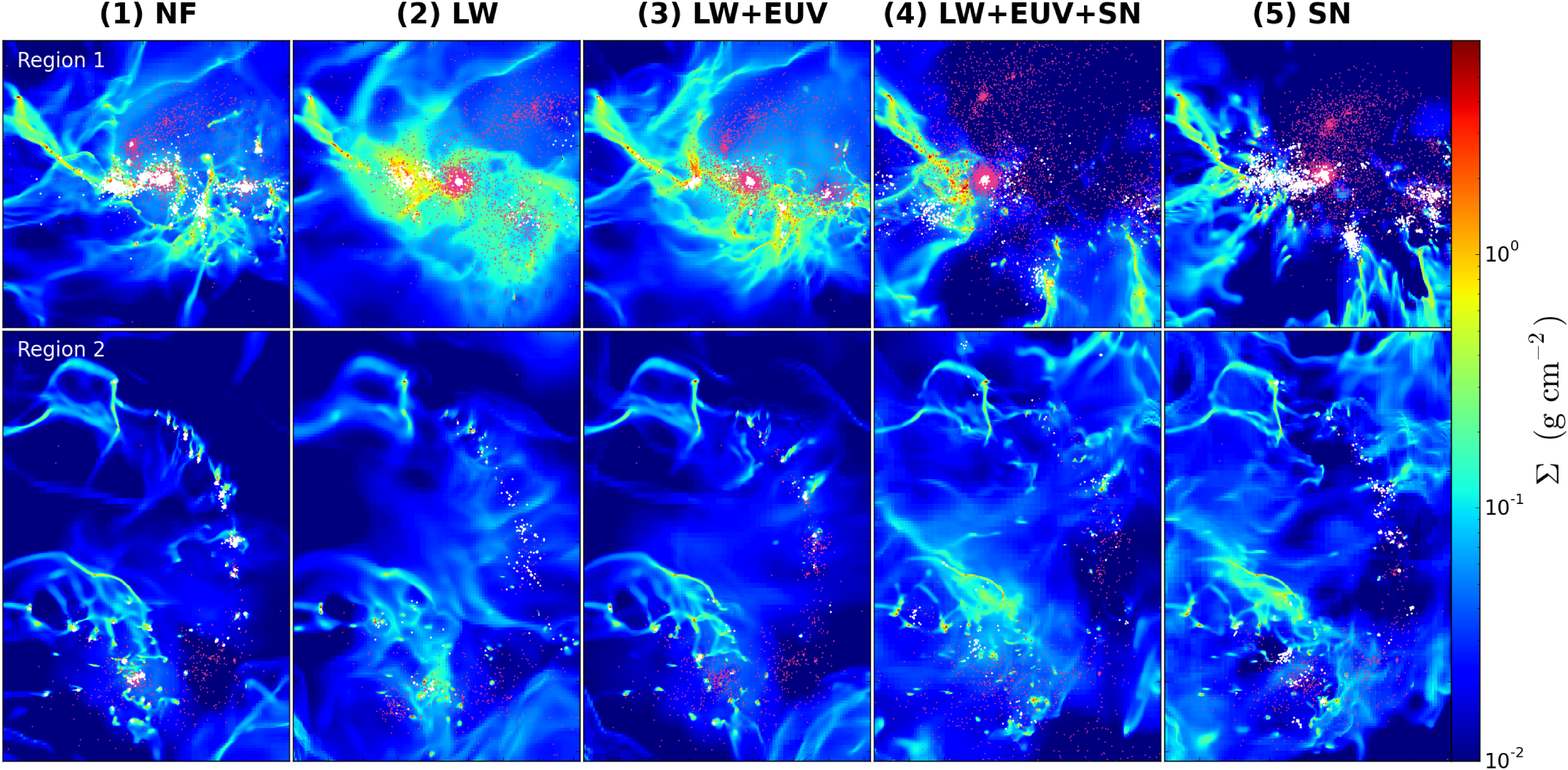} 
\end{array}$
\end{center}
\caption{ 
Zoom-in views of Region 1 (400~pc by 400~pc) ({\it top row}) and
Region 2 (400~pc by 600~pc) ({\it bottom row}) at $t=10$~Myr, showing
distribution of total gas mass surface density, $\Sigma_g$, along with
older stars formed before $t=5\:$~Myr (red points) and younger stars
formed after $t=5\:$~Myr (white points).
Columns from left to right show the results of simulations with: (1)
No Feedback (NF); (2) Only Dissociating Lyman-Werner Feedback (LW);
(3) Dissociating and Ionizing Feedback (LW+EUV); (4) Dissociating,
Ionizing and Supernova Feedback (LW+EUV+SN), which is the fiducial
model; (5) Only Supernova Feedback (SN).}\label{fig:f8}
\end{figure*}

To better visualize the locations of stars and gas and the effects of
different feedback implementations, in Figure~\ref{fig:f8} we show zoom-ins to
two regions at the time of 10~Myr. Region 1, a square patch of 400~pc
on a side, is centered in the disk midplane at ($x=+100$~pc,
$y=-200$~pc) relative to the center of the kpc box and contains the
most vigorous starburst activity. Region 2, which is 400~pc by 600~pc
in size and centered at ($x=0$~pc, $y=300$~pc), has a more moderate
level of star formation. These regions are the same as those
previously analyzed at 10~Myr in the MHD simulations of Van Loo et
al. (2015). Recall that each star particle represents a cluster or
sub-cluster with 100~$M_\odot$ that has formed during the last 10~Myr,
with the red color for the 0-5~Myr population (common to all runs) and
white color for those formed from 5 to 10~Myr and having varying
feedback properties. These figures give a dramatic visual illustration
of differences in the morphologies of molecular clouds and their star
formation activity that result from different implementations of
feedback physics. For example, in Region 1 much more massive star
clusters form in the no feedback run, while these are mostly dispersed
during the sequence LW, LW+EUV and LW+EUV+SN. The SN-only run has a
quite different distribution of young stars, which have had the most
destructive effects on their natal clouds.

To explore the differences in young stellar populations more
quantitatively, we identify the five most massive young star clusters
and/or associations in each simulation in Region 1, i.e., a starburst
region, selected to each be contained within $\sim 50$~pc$^{3}$-scale
volumes. For Run NF at 10 Myr, we find that the five most massive
clusters/associations (only counting stars formed after 5~Myr) have a
mean mass of $1.6\times 10^{5}\:M_{\odot}$. Introducing LW feedback
reduces this to $5.7\times 10^{4}\:M_{\odot}$; LW+EUV feedback further
reduces this to $5.1\times 10^{4}\:M_{\odot}$; and LW+EUV+SN feedback
to $1.2\times 10^{4}\:M_{\odot}$. On the other hand the SN only
feedback simulation produces five most massive clusters with mean mass
of $7.3\times 10^{4}\:M_{\odot}$. Thus the way in which feedback is
implemented can have dramatic impact on the clustering of star
formation.

\subsection{The Effect of Feedback on Star Formation Rates}
 
\begin{figure*}
\begin{center}$
\begin{array}{c} 
\includegraphics[width=6.1in]{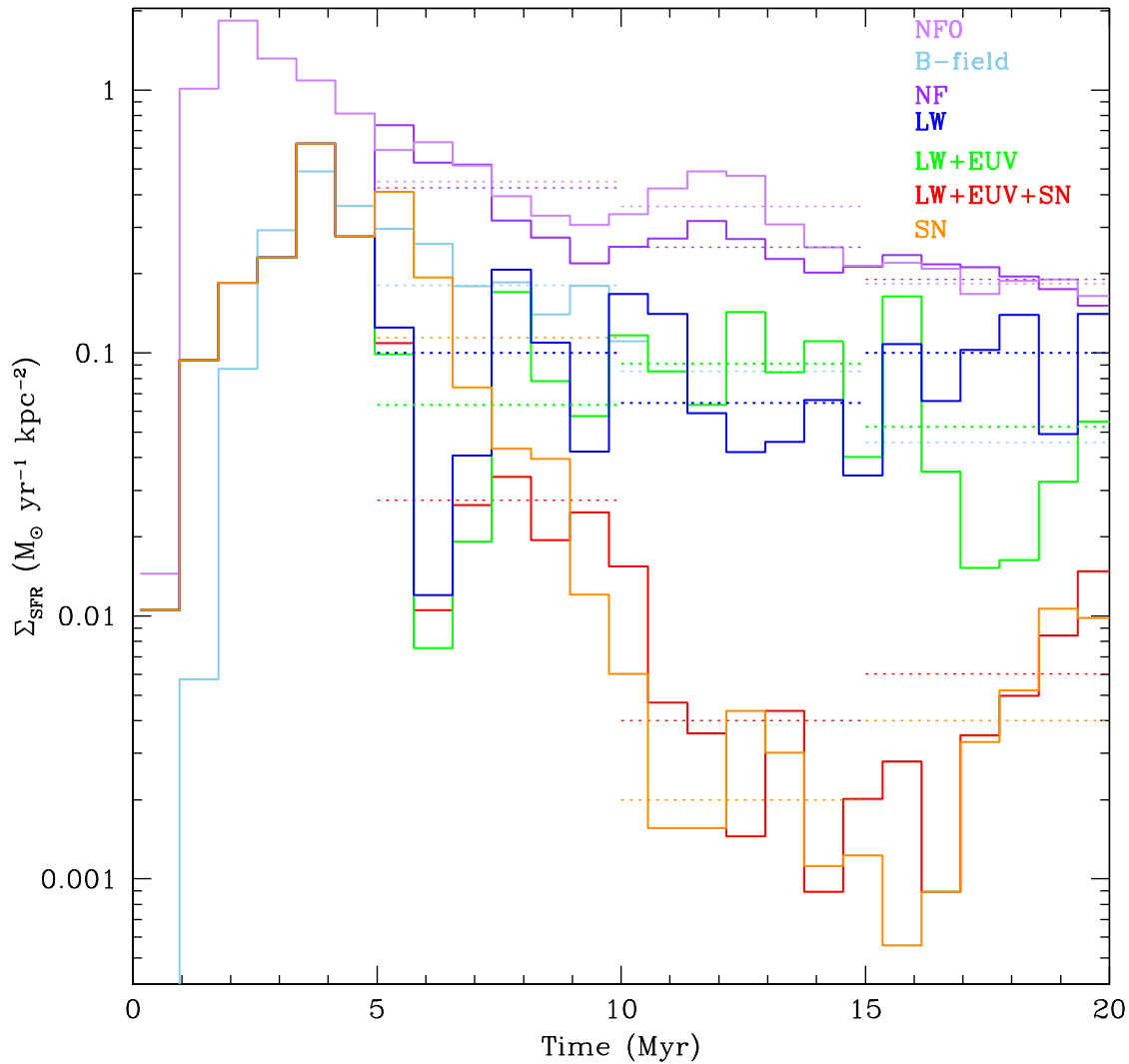}  \\
 \end{array}$
\end{center}
\caption{
Time evolution of star formation rate per unit area, $\Sigma_{\rm
  SFR}$, of the whole kpc patch of the disk for Run NF0 (light
purple), Run NF (purple), Run LW (blue), Run LW+EUV (green), Run
LW+EUV+SN (red) and Run SN (orange). The averages from 5-10~Myr,
10-15~Myr, and 15-20~Myr are indicated with horizontal dotted lines.
The no feedback run (Van Loo et al. 2015) with magnetic fields, which was carried out
to 10~Myr, is also shown (light blue), along with estimates of its
average SFRs from 10 to 20~Myr.
}\label{fig:f9}
\end{figure*}

We now examine the SFRs that occur during the simulations and the
effects of different feedback models. The time evolution of the SFR in
the whole kpc patch is shown in Figure~\ref{fig:f9}, averaging over time
intervals of just under 1~Myr. Here we also show the results of a run,
NF0, which has no feedback at all for the entire period from 0 to 20
Myr. This provides a baseline result against which the other
simulations can be compared. In addition, we plot the 0 to 10~Myr SFR
results of the MHD simulation of Van Loo et al. (2015) in which the
galactic disk patch is initially threaded with a uniform $B$-field of
strength of 10~$\rm \mu G$, along with an extrapolated SFR, assuming
exponential decay, from 10 to 20~Myr. As discussed by Van Loo et al. (2015),
magnetic fields help support the clouds against collapse and lead to
reductions in SFRs of factors of a few.

For the feedback simulation runs presented in this paper, all
implement the ``initial feedback model'' from 0 to 5 Myr, designed to
reach a realistic FUV intensity by this time. During this phase we see
suppression of the SFR by about a factor of ten in the first few Myr,
but then by smaller factors. After 5 Myr, the no feedback case
eventually converges towards a SFR approximately equal to the NF0 run
as the effects of stars formed in the initial feedback phase die
out. Introducing only Lyman-Werner feedback leads to factors of
several reduction in SFR compared to Run NF over most of the 20~Myr
evolution. However, the SFR now also shows much larger fluctuations,
which indicates the bursty nature of star formation and subsequent
feedback that is occuring in the simulated disk patch. We will thus
typically consider SFRs from the simulations averaged over 5~Myr
periods. The SFRs in the three 5~Myr period averages from 5 to 20~Myr
are fairly constant at a level of about $\Sigma_{\rm
  SFR}=0.1\:M_\odot\:{\rm yr^{-1}\:kpc^{-2}}$.

\begin{figure*}
\begin{center}$
\begin{array}{c} 
\includegraphics[width=6.75in]{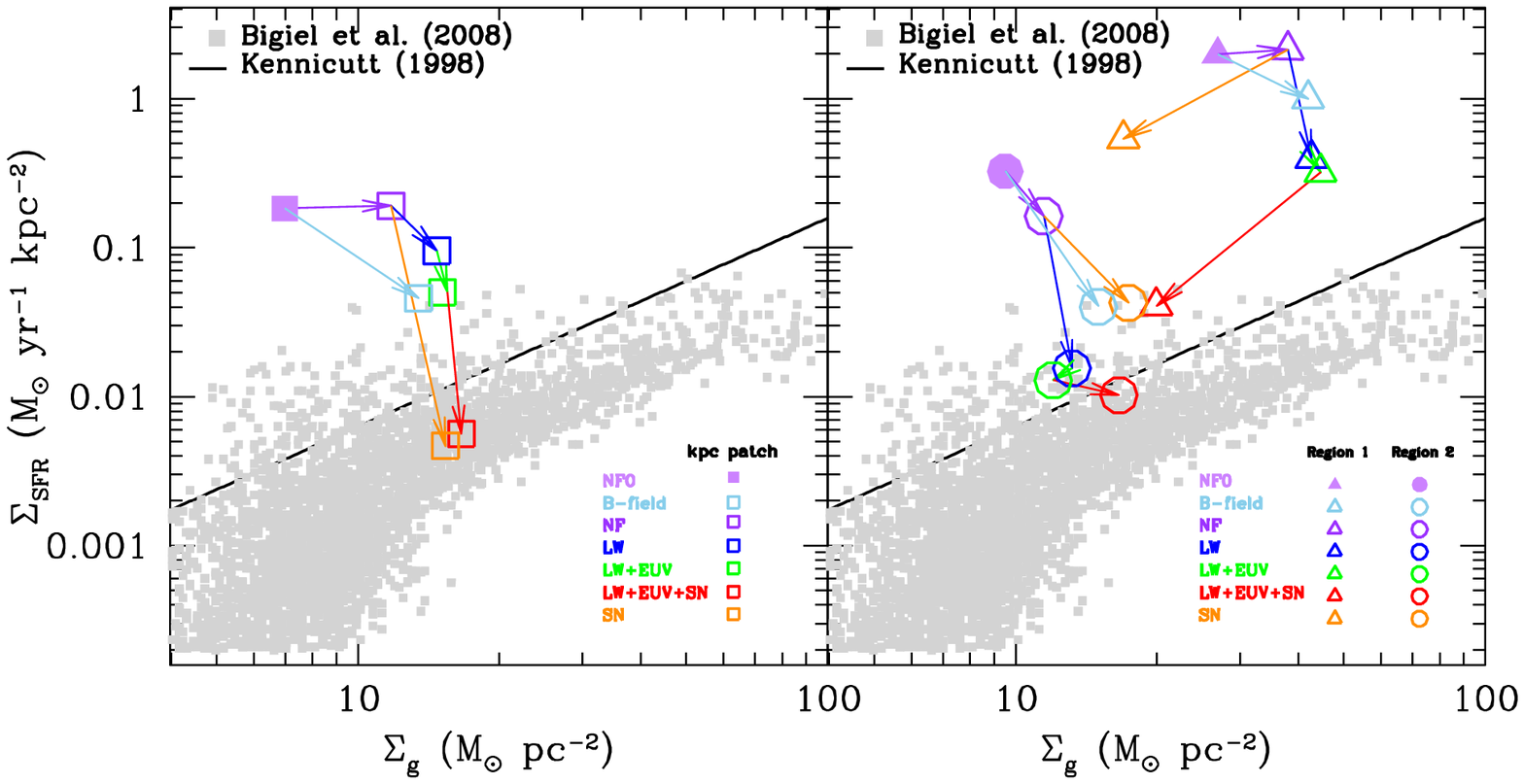}  \\
 \end{array}$
\end{center}
\vspace{-0.5in}
\vspace{-2.5in} 
\caption{
Star formation rate per unit area versus total gas mass surface
density. Left panel shows results for the whole kpc patch of the disk
with $\Sigma_g$ evaluated at 20~Myr and $\Sigma_{\rm SFR}$ being the
average from 15 to 20~Myr. The right panel shows results for Regions 1
and 2 with $\Sigma_g$ evaluated at 10~Myr and $\Sigma_{\rm SFR}$ being
the average from 5 to 10~Myr. The different simulation runs are
labeled in the legends (see text for description of the model
sequences). Observational data from Bigiel et al. (2008) are plotted as
gray squares (annuli of disk galaxies), while the empirical power law
fit to galactic and circumnuclear disk averages of Kennicutt (1998) is
shown with a black solid line.
%
}\label{fig:f10}
\end{figure*}

Continuing along the sequence of feedback models, the introduction of
ionizing feedback in Run LW+EUV leads to SFRs that, on average, are
only modestly smaller than Run LW, and sometimes can even be higher,
e.g., from 10 to 15 Myr. The final step to reaching the fiducial model
is the introduction of supernova feedback, i.e., Run LW+EUV+SN, which
leads to a significant further reduction in the SFRs. The global
evolution of the SFR appears to show a longer period systematic
fluctuation, decreasing from an initial relatively high value at 5 Myr
by almost a factor of 100 by the time around 15~Myr, before increasing
again to a value around $\Sigma_{\rm SFR}=0.01\:M_\odot\:{\rm
  yr^{-1}\:kpc^{-2}}$ by 20~Myr. However, the two 5~Myr period
averages from 10 to 20~Myr are relatively similar at about half this
value. Note that the absolute amount of reduction of $\Sigma_{\rm
  SFR}$ due to introduction of SN feedback in this sequence is
about the same as that caused by the earlier introduction of LW
feedback, relative to the NF case.

Finally, the SN-only feedback run shows an initially relatively high
SFR just after 5~Myr, which is understandable given the 3~Myr delay
associated with this feedback. However, the SFR then declines over the
next 10~Myr by about two orders of magnitude, before rising more
modestly again after 15~Myr. This SFR history is very similar to that
of the fiducial run with full feedback (LW+EUV+SN), which is
consistent with the very similar PDFs of gas mass surface density seen
in these models at late times (Fig.~\ref{fig:f7}).

Figure ~\ref{fig:f10}a shows the results of these simulations on a
``Kennicutt-Schmidt'' (KS) diagram of $\Sigma_{\rm SFR}$ versus
$\Sigma_{g}$, i.e., the total gas mass surface density. The left panel
shows conditions averaged over the whole kpc patch of the simulations,
with $\Sigma_{\rm SFR}$ evaluated as the average from 15 to 20~Myr,
i.e., when the statistical properties of the ISM appear to have mostly
stabilized (\S\ref{S:ism}). The value of $\Sigma_{g}$ is that observed at the
end of this period, i.e., at 20~Myr. This method of constructing the
KS diagram is similar to that adopted by observational studies, which
measure current gas properties but which evaluate SFRs from the
presence of massive star populations that are sensitive to the SFR
averaged over the last several Myr. The observational measurements of
Kennicutt (1998) and Bigiel et al.~(2008), the latter averaging over
scales of about 1~kpc, are plotted in Fig.~\ref{fig:f10}.

We have connected the results of the simulations with a sequence of
arrows that show the effect of adding particular physics. The pure no
feedback result, NF0, is in the upper left with a high SFR and, by
20~Myr, a much reduced gas content. The effect of magnetizing this
disk with a 10~$\rm \mu G$ field strength is shown by the light blue
arrow connecting to extrapolated results of Van Loo et al. (2015). The
purple arrow to Run NF illustrates the effect of the initial feedback
model implemented from 0 to 5 Myr, but then with feedback turned
off. The SFR near the end of the simulation is very similar to that of
NF0, even though the disk has retained more gas. Adding LW feedback
brings the SFR down by just over a factor of two and further raises
the gas mass surface density. Ionization has a more modest effect on
SFR, with remaining gas content (set mostly by earlier SFRs) hardly
affected. At the end of the sequence, the fiducial model of LW+EUV+SN
feedback brings the SFR down by about a factor of 10 at a slightly
larger gas content. We see that the fiducial model sits in the middle
of the distribution of observed SFRs of Bigiel et al. (2008). The
effect of the SN-only model is shown by the orange arrow from
NF. Overall it causes a similar reduction in SFR and increase in
retained gas content as the sequence to the fiducial model.

Figure~\ref{fig:f10}b shows similar results, but now for Region 1 (starburst
conditions) and Region 2 (normal disk conditions) (see Fig.~\ref{fig:f8}). Note,
these regions were originally defined at 10~Myr in the simulations of
Van Loo et al. (2013), and were also analyzed at this time in the MHD
simulations of Van Loo et al. (2015). By 20~Myr, the nature of these
regions, such as their gas content undergoes dramatic evolution. Thus
we also analyze the properties of these regions at 10~Myr, i.e., the
time at which $\Sigma_g$ is evaluated, with the SFR being the average
from 5 to 10~Myr. We note that this means conditions in the ISM in
these regions have not reached quasi-statistical equilibrium. However,
on these scales of only several hundred parsecs, such equilibrium is
in any case not likely to be achieved when averaging over 5~Myr
timescales. Observationally, at these scales of several hundred parsec
sided regions, a greater dispersion in the normalization of the
Kennicutt-Schmidt relation has been reported (Leroy et al., 2013).

Region 2, which is 600~pc by 400~pc in extent and with average
conditions quite similar to those of the kpc patch as a whole, shows
qualitatively similar behavior. However, LW feedback is seen to play a
relatively more important role here along the sequence to the fiducial
model of LW+EUV+SN. The fiducial model rests at a location slightly
elevated in its normalization compared to the median of the data,
although still within the range of the observed scatter in the KS
relation. Since Region 2 is centered on a GMC complex, this slight
enhancement is also consistent with theoretical expectations, given
the clustered nature of star formation in disk galaxies. In Region 2
the $B$-field model also leads to a significant reduction in SFR,
although not as much as that due to the fiducial feedback model. The
SN-only feedback model is also not as effective as the full radiative
feedback plus SN model in this case.

Region 1, centered on a very dense grouping of GMCs that are also
interacting, shows very high SFR surface densities. The NF0 and NF
models have $\Sigma_{\rm SFR}\simeq2\:M_\odot\:{\rm
  yr^{-1}\:kpc^{-2}}$, with the latter retaining a gas mass surface
density of about $40\:M_\odot\:{\rm pc}^{-2}$ at 10~Myr. LW feedback
brings the SFR down by about a factor of five, with addition of
ionization causing only a minor reduction beyond this. The final
LW+EUV+SN model has its SFR reduced further by a factor of 10 to
$\Sigma_{\rm SFR}\simeq0.04\:M_\odot\:{\rm
  yr^{-1}\:kpc^{-2}}$. However, note the largest reduction of SFR in
absolute terms resulted at the stage of introduction of FUV $\rm H_2$
dissociating feedback. The introduction of SN causes leads to a
dramatic reduction in $\Sigma_g$, i.e., by about a factor of three,
which is caused by expulsion of gas from Region 1 to its immediate
surroundings. This illustrates the inherent difficulties of measuring
the KS relation on these small scales of several hundred
parsecs. Still, the final location of Region 1 in the KS diagram for
the fiducial model is at a normalization that is consistent with the
upper envelope of the observational data, as might be expected for a
starburst region.

Finally, we note that introduction of magnetic field support affects
the SFR of Region 1 only by about a factor of two, a smaller factor
than in Region 1 and in the kpc patch as a whole. This is because the
dense GMCs in Region 1 would need stronger $B$-field strengths to
resist collapse to the same level as the lower density GMCs in the
other parts of disk. SN-only feedback reduces SFRs by factors of a few
and, like the case of LW+EUV+SN, expels most of the gas from the
region. The final location of the SN-only model in the KS diagram is
at a level that is about 10 times higher than the upper envelope of
the observed data.

\section{Discussion and Conclusions}\label{S:discussion}

We have found that the combined effects of dissociating, ionizing, and
supernova feedback have a dramatic influence on the structure and star
formation activity of the interstellar medium of a disk galaxy. The
relative mass and volume fractions of different phases of the ISM
approach quasi-static values after 20~Myr of evolution of the
simulated kpc patch of the disk. The detailed structure of the ISM,
especially dense, molecular clouds, and the spatial distribution of
young stellar populations varies significantly with how feedback is
implemented, as we have also seen in comparison to a supernova only
feedback model.

Of the forms of feedback we have investigated in the sequence of
  building our fiducial model, it is the introduction of FUV
radiation that dissociates $\rm H_2$ molecules and heats the gas,
which typically plays the largest role in reducing absolute SFRs
from those seen in simulations with no feedback from the newly forming
stars.  These reduction factors are similar to those seen in models
including magnetic support from an ISM magnetized at a level of
$10\:{\rm \mu G}$. However, the full combination of radiative and
mechanical feedback from supernovae of our fiducial model is needed to
bring these overall star formation rates down to values similar
to those observed in real disk systems. Averaged on kpc scales, the
SFRs resulting from the SN-only feedback models are very similar to
those achieved in our fiducial simulation, even though these involve
large differences in the chemical and temperature structures of the
gas and in the spatial distribution of young stars. Also, in smaller
scale regions centered on GMC complexes, there are significant
differences in the SFRs between the SN-only case and the full feedback
model. These results illustrate the importance of considering FUV
feedback in understanding the star formation activity of galactic disk
systems. Similar conclusions have been reached by Peters et
  al. (2017) in their lower (4~pc) resolution studies of a smaller
  (0.5 kpc by 0.5~kpc) disk patch, which builds on the previous study
  of a stellar wind and supernova feedback model by Gatto et
  al. (2017).

The simulations presented in this paper of course have their
limitations and there is scope for improvement, which may influence
the final results. For example, it will be necessary to combine these
feedback models with magnetized ISM simulations. Additional feedback
effects such as protostellar outflows from forming stars, stellar
winds and radiation pressure will also need to be considered. The
momentum input from these sources can be comparable to that from
supernovae, although in the case of outflows and winds much of this
momentum may be lost in oppositely directed flows from a distributed
population of stars. Also, initial test simulations including
radiation pressure by Rosdahl et al., in prep., indicate that it plays
a relatively minor role on the scale of this kpc patch. On smaller
scales radiation pressure has been studied by, e.g., Krumholz \&
Matzner (2009) who find it can be important feedback in modifying HII
region dynamics around massive star clusters. The contribution of
relativistic cosmic rays, which are known to have a similar energy
density to magnetic and thermal pressures in the local Galactic ISM,
will also need to be included. To constrain these more complex models
and their particular parameter choices against observed systems, we
will also need to examine more detailed statistical metrics, e.g., of
structural, kinematic and chemical properties of the ISM, in addition
to SFR activity. Improved resolution down to the core scale of
individual star formation is also highly desirable, especially for
resolution of the early stages of ionizing feedback, which will then
also necessitate a fully stochastic treatment of the initial stellar
mass function of massive stars. On the other hand, the region we have
simulated is relatively limited in spatial extent, which may lead to
stochastic effects related to the particular GMCs that are included in
the volume. Future work should also aim to expand the number of
different realizations simulated, the size of the simulated volume and
also explore the effects of a range of galactic environments.

%

\clearpage

\appendix
\section{Appendix: Additional Simulation Results}
\counterwithin{figure}{section}

\begin{figure*}[h!]
\begin{center}$
\begin{array}{c} 
\includegraphics[width=6.7in]{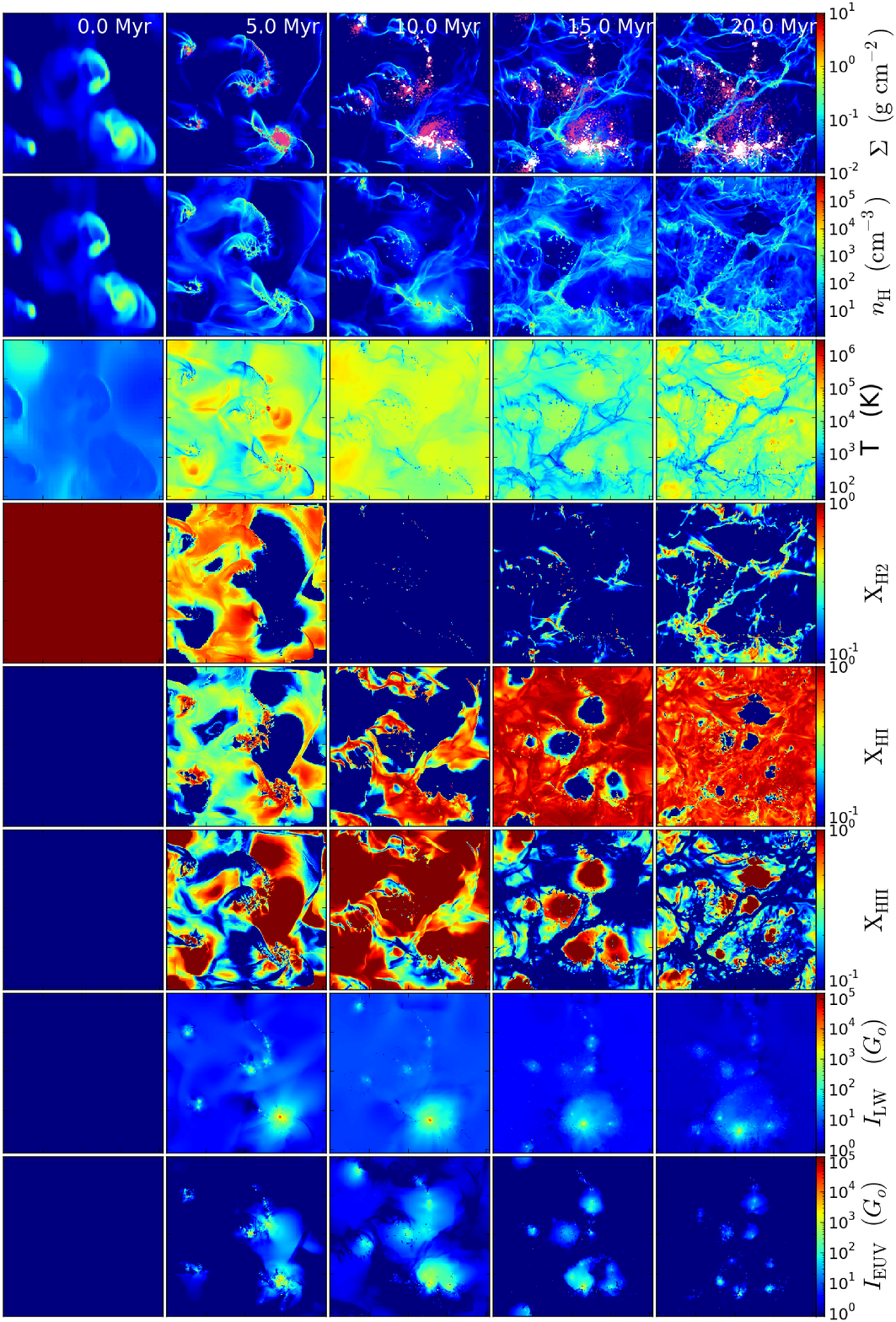} 
\end{array}$
\end{center}
\caption{ 
Same as Fig.~\ref{fig:f1}, but for the No Feedback simulation (Run NF).
}\label{fig:A1}
\end{figure*}
\clearpage
\begin{figure*}
\begin{center}$
\begin{array}{c} 
\includegraphics[width=6.7in]{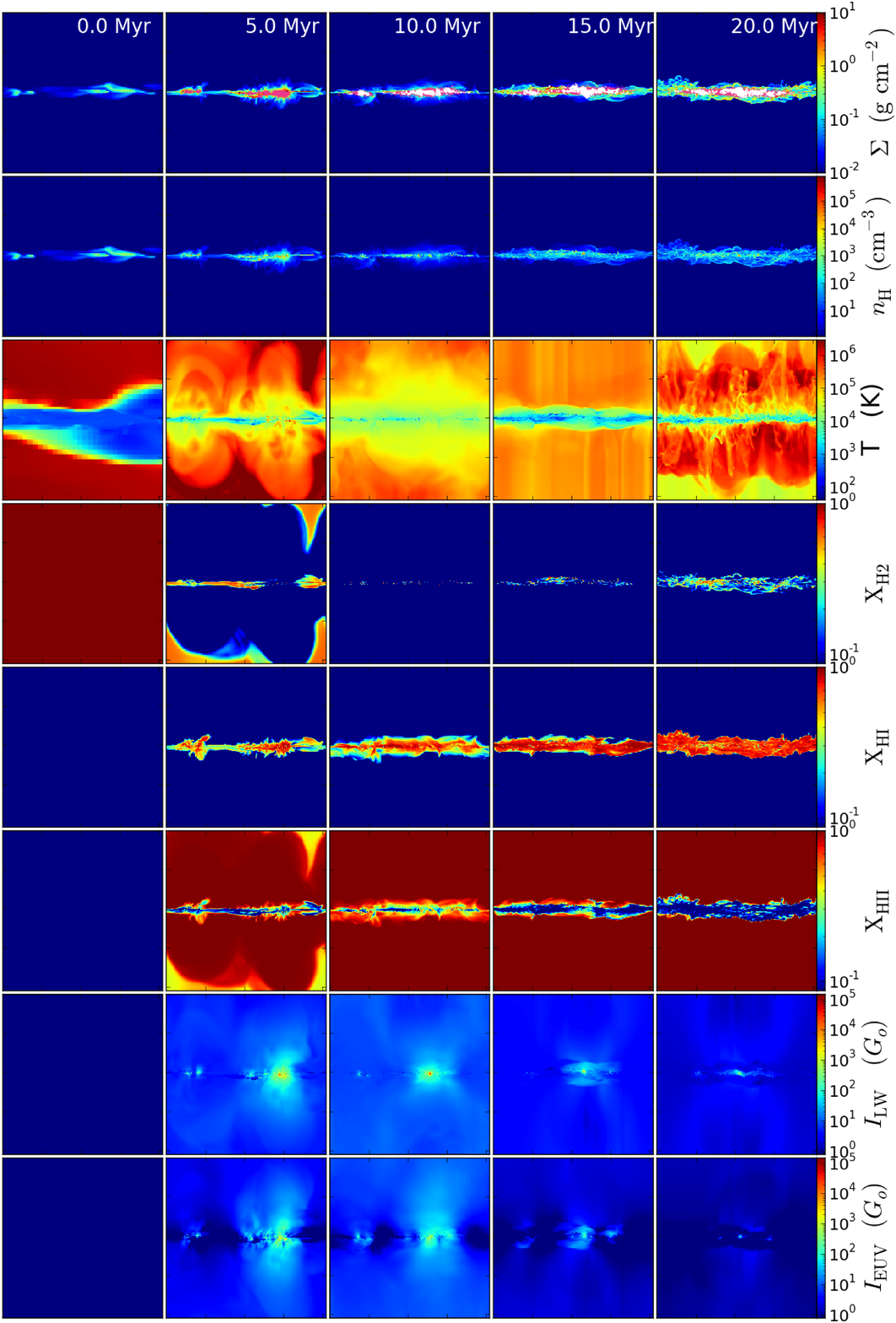} 
\end{array}$
\end{center}
\caption{ 
Same as Fig.~\ref{fig:f1}, but for the No Feedback simulation (Run NF).
}\label{fig:UVz}
\end{figure*}
\clearpage
\begin{figure*}
\begin{center}$
\begin{array}{c} 
\includegraphics[width=6.7in]{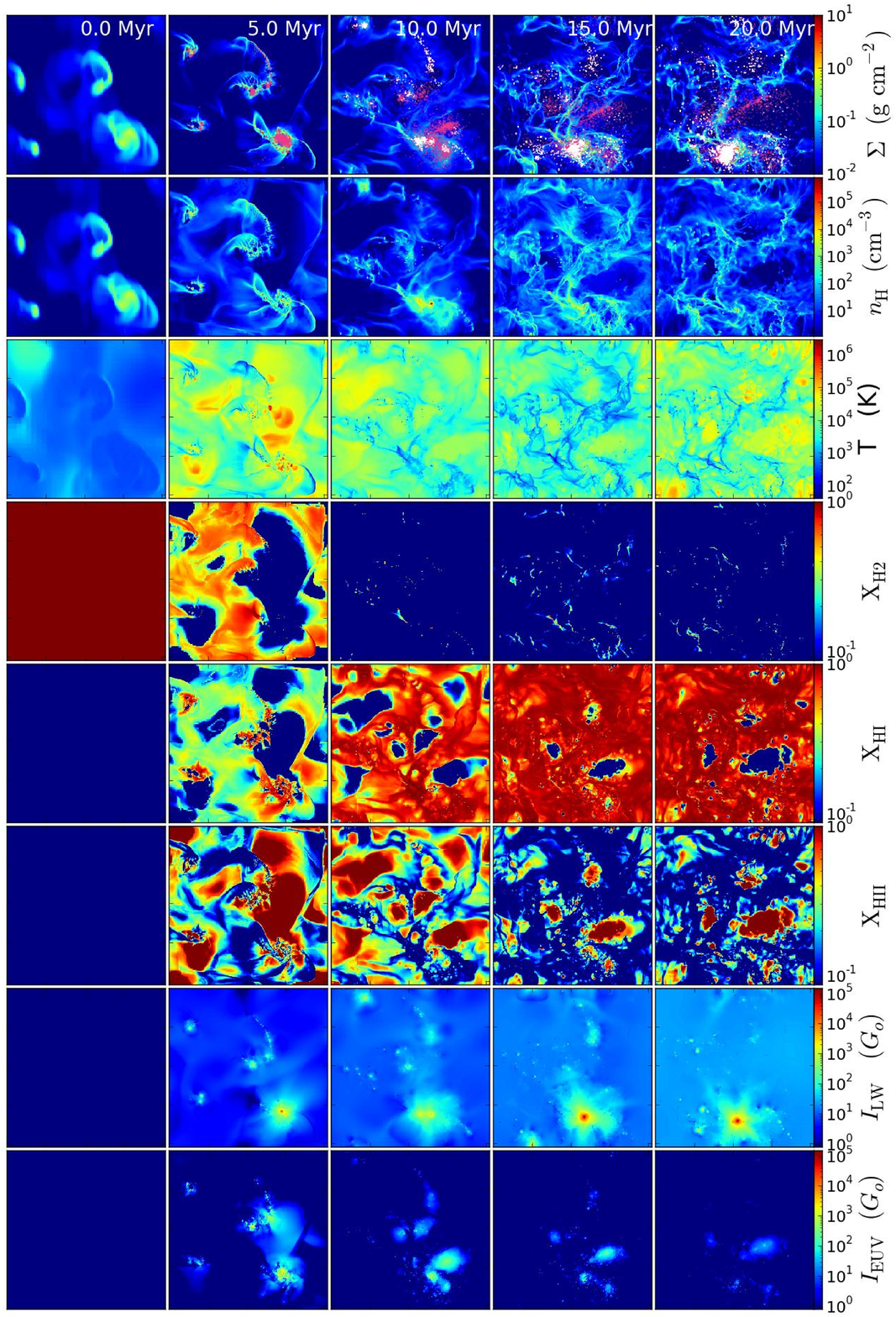} 
\end{array}$
\end{center}
\caption{
Same as Fig.~\ref{fig:f1}, but for the Lyman-Werner only simulation (Run LW).
}\label{fig:UVz}
\end{figure*}
\clearpage
\begin{figure*}
\begin{center}$
\begin{array}{c} 
\includegraphics[width=6.7in]{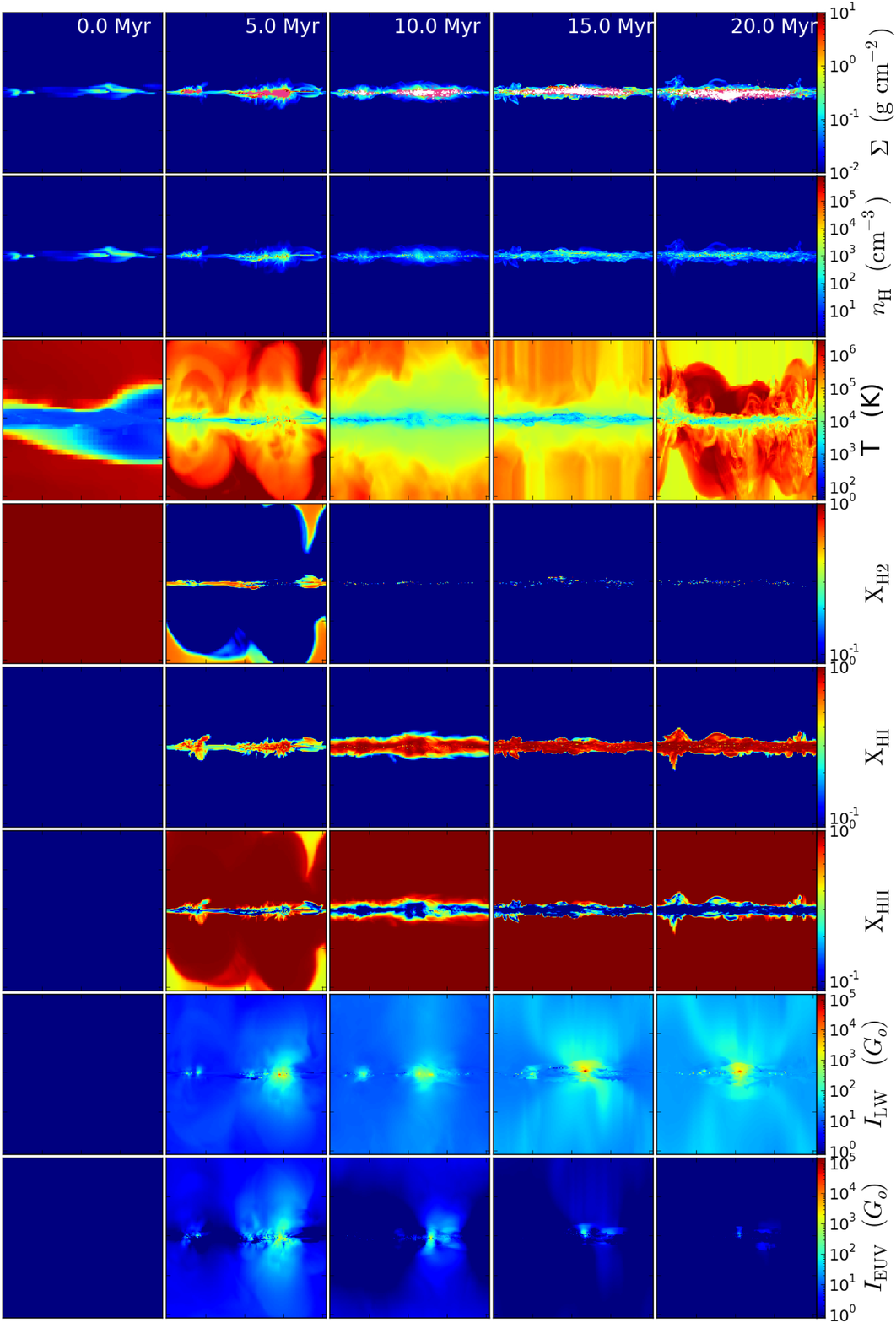} 

\end{array}$
\end{center}
\caption{ 
Same as Fig.~\ref{fig:f1}, but for the Lyman-Werner only simulation (Run LW).
}\label{fig:UVz}
\end{figure*}
\clearpage

\begin{figure*}
\begin{center}$
\begin{array}{c} 
\includegraphics[width=6.7in]{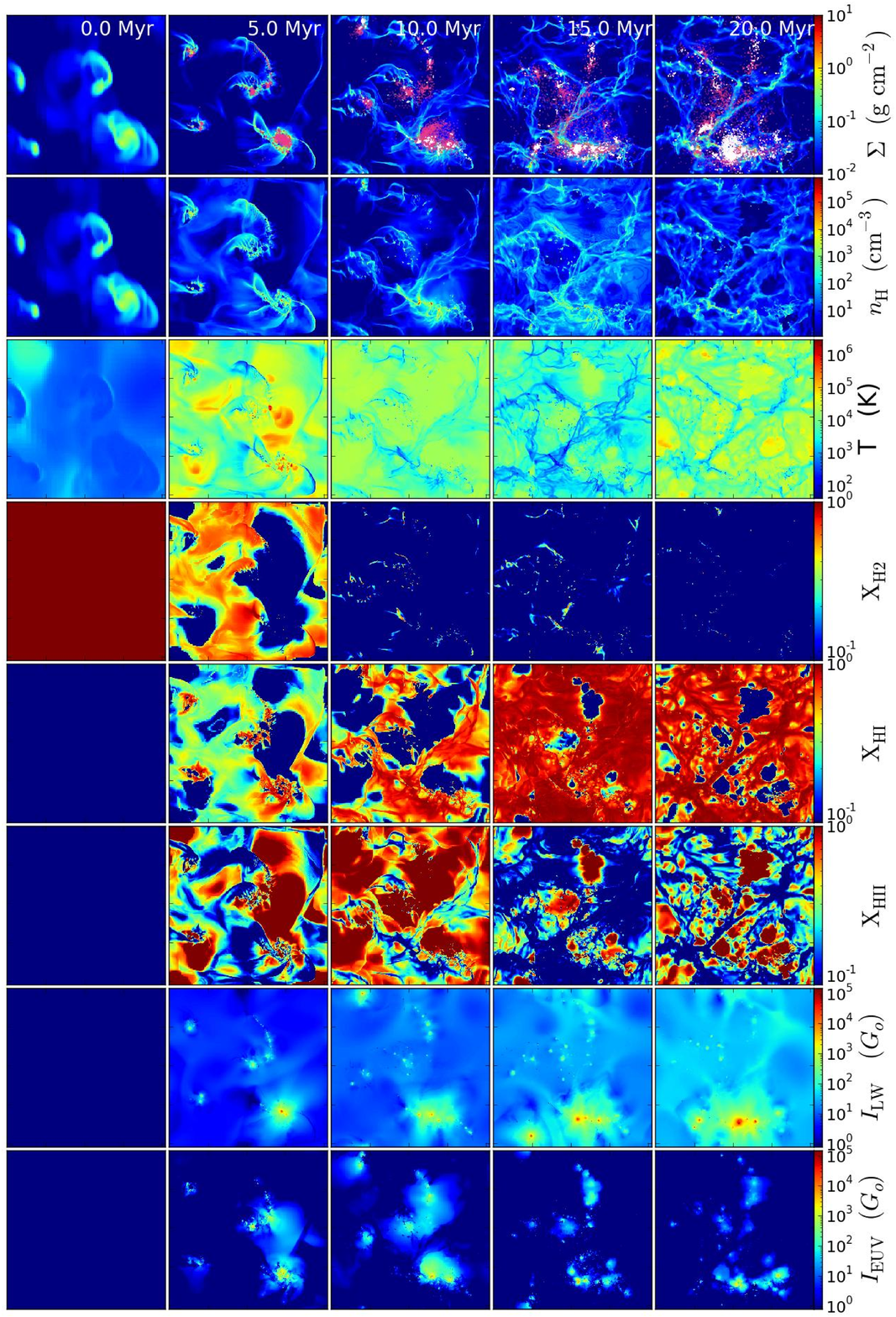} 

\end{array}$
\end{center}
\caption{ 
Same as Fig.~\ref{fig:f1}, but for the Lyman-Werner and Ionization simulation (Run LW+EUV).
}\label{fig:UVz}
\end{figure*}
\clearpage
\begin{figure*}
\begin{center}$
\begin{array}{c} 
\includegraphics[width=6.7in]{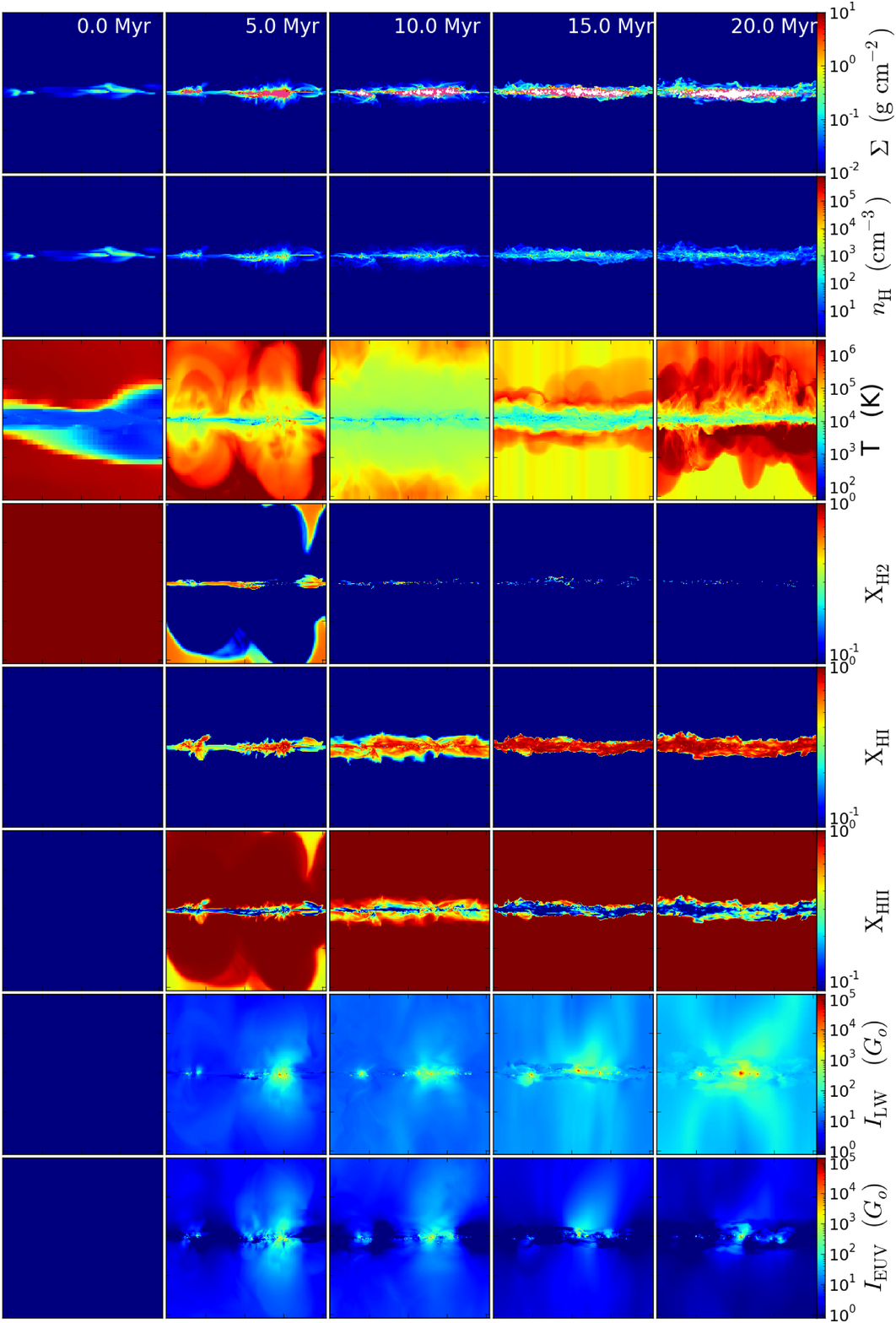} 

\end{array}$
\end{center}
\caption{
Same as Fig.~\ref{fig:f1}, but for the Lyman-Werner and Ionization simulation (Run LW+EUV).
}\label{fig:UVz}
\end{figure*}

\clearpage

\begin{figure*}
\begin{center}$
\begin{array}{c} 
\includegraphics[width=6.7in]{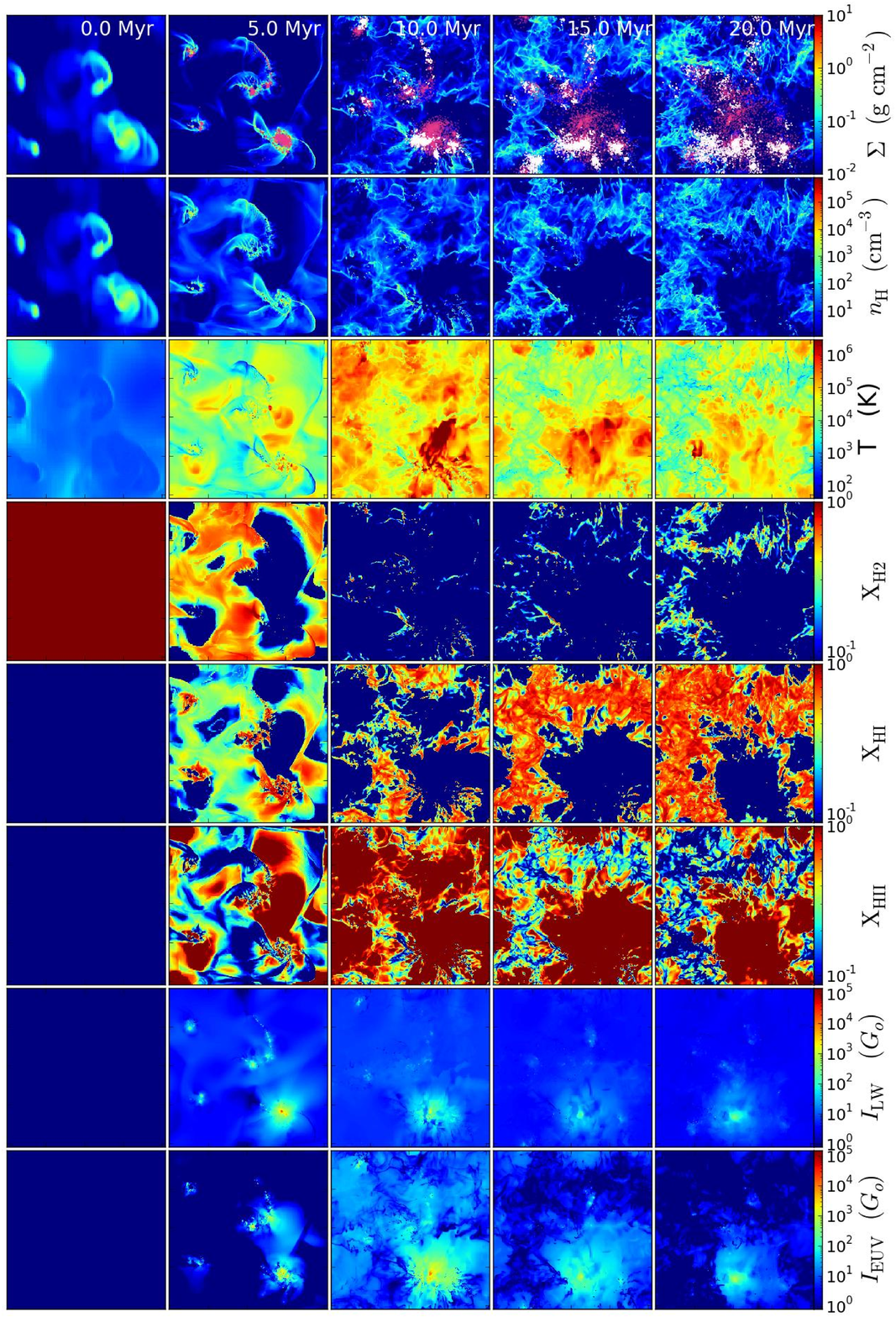} 

\end{array}$
\end{center}
\caption{ 
Same as Fig.~\ref{fig:f1}, but for the Supernova only simulation (Run SN).
}\label{fig:SNz}
\end{figure*}
\clearpage
\begin{figure*}
\begin{center}$
\begin{array}{c} 
\includegraphics[width=6.7in]{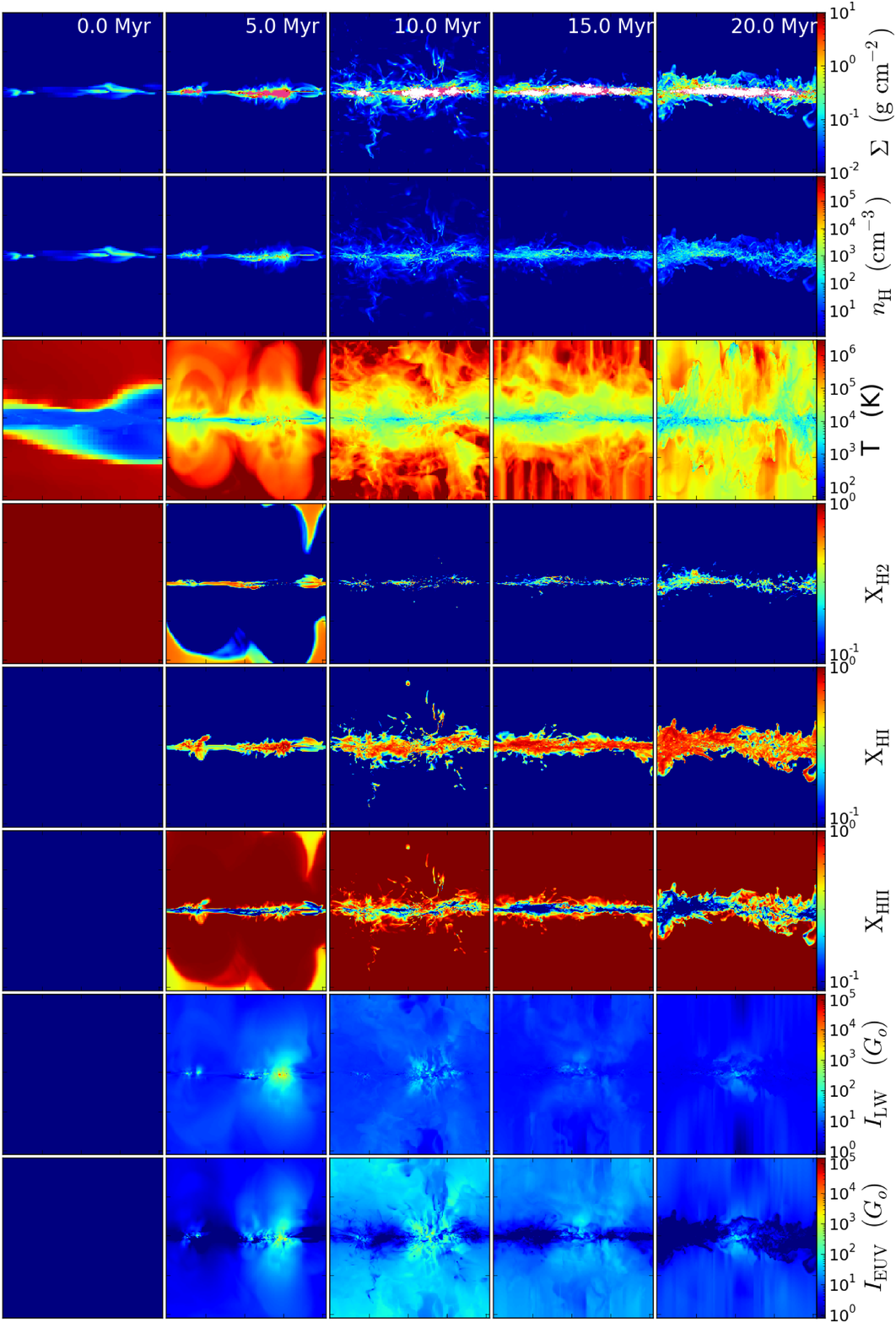} 

\end{array}$
\end{center}
\caption{ 
Same as Fig.~\ref{fig:f1}, but for the Supernova only simulation (Run SN).
}\label{fig:A8}
\end{figure*}

\clearpage



\clearpage




\clearpage




\end{document}